# From Conservatism to Innovation: The Sequential and Iterative Process of Smart Livestock Technology Adoption in Japanese Small-Farm Systems

Takumi Ohashi[1*, 2], Miki Saijo[1], Kento Suzuki[3], Shinsuke Arafuka[3]

[1]Tokyo Institute of Technology, 2-12-1 Ookayama Meguro-ku, Tokyo 152-8550, Japan
[*]ohashi.t.af@m.titech.ac.jp
[2]Chulalongkorn University, Phayathai Road, Pathumwan, Bangkok 10330, Thailand
[3]Eco-Pork Co. Ltd., 3-21-7 Kandanishikicho Chiyoda-ku, Tokyo 101-0054, Japan

## Abstract

As global demand for animal products is projected to increase significantly by 2050, driven by population growth and increased incomes, smart livestock technologies are essential for improving efficiency, animal welfare, and environmental sustainability. Conducted within the unique agricultural context of Japan, characterized by small-scale, family-run farms and strong government protection policies, our study builds upon traditional theoretical frameworks that often oversimplify farmers' decision-making processes. By employing a scoping review, expert interviews, and a Modified Grounded Theory Approach, our research uncovers the intricate interplay between individual farmer values, farm management policies, social relations, agricultural policies, and livestock industry trends. We particularly highlight the unique dynamics within family-owned businesses, noting the tension between an "advanced management mindset" and "conservatism." Our study reveals that technology adoption is a sequential and iterative process, influenced by technology availability, farmers' digital literacy, technology implementation support, and observable technology impacts on animal health and productivity. These insights highlight the need for tailored support mechanisms and policies to enhance technology uptake, thereby promoting sustainable and efficient livestock production system.

# 1. Introduction

Global livestock demand is projected to increase by 66% by 2050 compared to 2005/2007 levels (Alexandratos and Bruinsma, 2022), driven by growing populations and rising incomes. This significant increase in demand highlights the need for sustainable and efficient livestock production systems to ensure food security and reduce environmental impact (Pelletier and Tyedmers, 2010; Thornton, 2010). New smart livestock technologies, such as precision feeding, automated milking systems, and wearable animal health monitoring devices, have the potential to improve efficiency, animal welfare, and environmental sustainability (Eastwood et al., 2012; Li et al., 2021; Takizawa et al., 2022; Wolfert et al., 2017).

However, the full potential of these technologies can only be realized if farmers adopt them. While theoretical frameworks like the Technology Acceptance Model (TAM) and the Function of Innovation System Framework have been employed to examine technology adoption, they may not fully capture the complexities influencing farmers' decisions to adopt smart livestock technologies (Kebebe, 2019; Michels et al., 2019). This gap in understanding requires a more detailed exploration of the factors that influence technology adoption, especially within specific cultural and industrial contexts.

Japan, known for its technological advancements and unique socio-cultural factors impacting farming practices, provides a relevant context for this study. Despite its technological capabilities, the adoption of smart livestock technologies in Japan has been slower compared to other developed countries. This discrepancy highlights the need to understand the specific barriers and drivers within the Japanese agricultural context.

For this study, we define smart livestock technology based on the guidelines provided by the Japanese Ministry of Agriculture, Forestry and Fisheries (MAFF, 2019). This includes technologies for sensing and monitoring biological functions and breeding environments, AI-based data utilization, automated operations, and business data management. A detailed definition is provided in the Materials and Method section.

To address this gap, this study uses a three-pronged methodology, including a scoping review, expert interviews, and a Modified Grounded Theory Approach (M-GTA), to explore the landscape of Japan's agricultural sector. The research aims to answer the following question: what are the factors that influence the adoption of smart livestock technologies among Japanese farmers, including both barriers and facilitators?

By answering this question, this research aims to develop a comprehensive understanding of the factors that influence smart livestock technology adoption in Japan. Our findings will provide valuable insights for policymaking and stakeholders, helping to create effective strategies and interventions to promote the adoption of sustainable and efficient livestock farming system. This study emphasizes the importance of understanding technology adoption within specific cultural and industrial contexts, contributing to the development of relevant policies and strategies.

## 2. Case Study Context

According to Geospatial Information Authority of Japan (GSI), Japan is an archipelago in East Asia, located approximately between 20 and 46 degrees north latitude and 122 and 154 degrees east longitude (GSI, 2019), covering about 378,000 km² (GSI, 2024). The country features diverse landscapes, including plains, hilly areas, mountainous regions, and remote islands, with mountainous areas comprising about 75% of the total land area (GSI, 1990). Japan has a total population of 126.146 million, with approximately 28.6% aged 65 or older (Statistics Bureau, 2021), making it the country with one of the highest aging rate in the world.

Japanese agriculture is characterized by its small-scale management structure, geographical constraints, historical background, cultural factors, and government protection policies, which have hindered large-scale farming compared to other developed countries (Su et al., 2018). According to FAOSTAT, the average farm size per farmer in Japan is 2.2 hectares, significantly smaller compared to the average size of about 42 hectares in France, 160 hectares in the United States, and 1143 hectares in Australia (calculated by the authors from FAOSTAT by FAO (n.d.)).

Japan's predominantly mountainous terrain results in narrow and dispersed farmland, with many hilly and mountainous regions spread across the country (Hoshino, 2001; Uematsu et al., 2010). In contrast, many European countries with prominent agricultural sectors have large areas of farmland, and countries like the United States, Canada, Australia, and France have vast agricultural lands where large-scale commercial farming is common (Bokusheva and Kimura, 2016). To fully understand the differences in technological adoption processes in agriculture, it is crucial to consider both geographical constraints and historical contexts.

Before World War II, Japanese farmers adapted to an industrializing economy by developing small-scale agricultural machinery, essential for the survival of small farms post-war. Pre-war technological advancements supported a "pluriactivity" strategy, enabling farmers to engage in both agricultural and non-agricultural income activities. For instance, in Okayama Prefecture, farmers modified machinery to suit their needs, leading to the development of the power tiller, which became central to post-war mechanization efforts (Francks, 1996). After World War II, Japanese agriculture underwent significant transformation. Under the occupation policies of the General Headquarters (GHQ), land reforms dismantled the feudal landlord system, redistributing land to tenant farmers (Noda, 1997). This resulted in the emergence of small-scale owner-farmers, with family-run small-scale farming becoming the mainstream. Additionally, the protective policies of the Japanese government and the activities of Japan Agricultural Cooperatives (JA) economically supported small-scale farmers (Godo, 2002). Currently, one of the most severe challenges facing Japanese agriculture is the aging agricultural workforce. Over 50% of agricultural workers are aged 65 or older (calculated by the authors from FAOSTAT by FAO (n.d.))., and this demographic trend necessitates the introduction of labor-saving technologies. While similar challenges are seen in other developed countries, the severity is pronounced in Japan due to its demographic structure. Particularly in mountainous areas, maintaining agriculture is challenging due to a lack of successors, aging farmers, and population outflow, making it difficult to immediately address labor shortages (MAFF, 2024).

In this context, the introduction of smart technology has become an urgent issue. Additionally, with increasing awareness of environmental issues, there is growing momentum to adopt technologies that reduce excessive use of fertilizers and pesticides while maintaining high productivity with minimal environmental impact. Supported by government subsidies and assistance programs, the introduction of new technologies has started to progress. Future agricultural policies related to smart technology are moving towards supporting a highly productive food supply system that can maintain production levels even under population decline. Efforts are being made to promote the development of smart technologies through industry-academia-government collaboration and to transform production, distribution, and sales systems to respond to smart technologies as an integrated supply chain (MAFF, 2024).

Therefore, as discussed, Japan's agriculture is characterized by small-scale farms, diverse natural environments, and cultural practices that vary regionally. Nationally, Japan faces uniform challenges such as population decline and aging, which agricultural policies aim to address. Considering the diversity at the farm level and the uniformity at the policy level, understanding the adoption of smart technologies in livestock farming requires a comprehensive view of the system from the micro-level of individual farms to the macro-level of national policies.

# 3. Materials and Method

This study employed a three-pronged approach, consisting of a scoping review, expert interviews, and an M-GTA to elucidate the process of smart livestock technology adoption by livestock farmers. The overall framework of our research is based on the Constructivist Grounded Theory (CGT) approach (Charmaz, 2014), which provides the theoretical foundation for our methodology.

CGT differs from the traditional Grounded Theory Approach in that it recognizes theories as co-constructed through interactions between the researcher and participants. This approach incorporates the researcher's background and existing literature to frame research questions and contextualize findings. Therefore, the scoping review in our study serves several purposes aligned with CGT principles: it provides a comprehensive understanding of existing knowledge, identifies key themes related to smart livestock technology adoption, and helps in developing relevant interview questions. However, our primary analysis remains firmly grounded in the data collected from expert interviews. To analyze this data, we adopted the M-GTA, as proposed by Kinoshita (2020). M-GTA is designed for the development of domain-specific, substantive theories rather than broad-based formal theories. This method aligns with CGT principles by emphasizing the contextual and co-constructed nature of theories. It involves systematic coding and categorization of data to generate theories that are firmly grounded in the specific context of the study.

## 3.1. Scoping review

The scoping review aimed to identify relevant articles that examined factors influencing smart livestock technology adoption behavior. A comprehensive search executed in accordance with the PRISMA-ScR method (Tricco et al., 2018) and conducted on December 21, 2021, utilizing the Web of Science All Databases, which includes Web of Science Core Collection, BIOSIS Citation Index, Current Contents Connect, Data Citation Index, Derwent Innovations Index, KCI - Korean Journal Database, MEDLINE, Russian Science Citation Index, SciELO Citation Index, and Zoological Record. The search terms for the literature review were selected to capture a broad range of studies related to smart livestock technology adoption. The primary term 'livestock' was chosen to include various types of animal agriculture. The search query used was as follows: TS = (technology) AND TI = (acceptance OR adoption OR uptake) AND TS = (livestock).

To maintain rigor and relevance in the scoping review process, we established specific inclusion and exclusion criteria for the selection of articles to be included in the analysis. Table 1 shows the inclusion and exclusion criteria used.

**Table 1 Inclusion and exclusion criteria for scoping review.**

| Inclusion criteria | Exclusion criteria |
|---|---|
| Peer-reviewed, open-access, full-text articles | Publications limited to review papers, conference proceedings, reports, or abstracts only |
| Articles published in the English language | Articles published in languages other than English |
| Studies exploring factors that influence the adoption of information technology (IT) and precision agriculture technologies among livestock farmers | Studies examining factors unrelated to IT technologies, such as institutional or organizational affiliations, in relation to acceptance |
| Research focusing primarily on livestock farming | Research centered on farmers primarily engaged in the production of non-livestock agricultural commodities |
| Smart technologies pertinent to livestock production processes, defined by Japanese Ministry of Agriculture, Forestry and Fisheries (MAFF, 2019) | Other technologies associated with marketing or consumption stages |

The definition of smart technology used in this study was based on the definition of the Japanese Ministry of Agriculture, Forestry and Fisheries, which includes the following components: sensing and monitoring technology to provide data on biological functions (such as reproductive function, nutrition, and health status) and the breeding environment; AI-based utilization of biological data; AI-based utilization of breeding environment data; technology to automate operations and reduce labor through the introduction of automated driving robots; and technology to manage business data, such as analyzing the current state of management, making plans, and monitoring progress (MAFF, 2019).

By adhering to these refined inclusion and exclusion criteria, we ensured that the articles selected for the scoping review were both relevant and contributed to our research objective of understanding the factors affecting the adoption of smart livestock technologies by livestock farmers.

Data extraction involved identifying and recording key information from each article, including the country of study, livestock breed, technologies, methods, theoretical frameworks, and factors influencing technology adoption. This study employed an open coding approach to extract factors associated with technology adoption by livestock farmers. To elucidate the factors, the first author executed open coding by thoroughly reading each article’s title, abstract, and full text. The generated codes were subsequently reviewed, synthesized, and merged into second-order codes, referred to as extracted factors in this study.

In this study, a single reviewer (first author) with extensive experience in conducting scoping reviews across multiple disciplines was responsible for the screening process. The reviewer's experience is supported by their previous work in scoping reviews, which can be cited as evidence of their expertise in this area (Ohashi et al., 2022; Takagi et al., 2023; Zallio et al., 2023; Zallio and Ohashi, 2022). The use of a single experienced reviewer was deemed appropriate for this rapid review, as it has been suggested that single-reviewer screening can be a suitable methodological shortcut for rapid reviews when conducted by an experienced reviewer (Waffenschmidt et al., 2019).

Despite the comprehensive search strategy and rigorous inclusion criteria, our review faced certain limitations. The use of the term 'livestock' might have limited our search results, potentially excluding studies focusing on specific types of livestock such as dairy, sheep, pigs, and poultry. To mitigate this limitation, we conducted expert interviews with key stakeholders in the agricultural sector, which helped uncover additional factors influencing technology adoption that may not have been evident in the literature. Additionally, the single-reviewer screening approach, while efficient, may have inherent limitations. Future research could benefit from multiple reviewers to enhance the robustness of the screening process. Lastly, the decision to include only open-access articles published in English excluded several potentially relevant studies. Non-English and non-open-access articles might have provided additional valuable insights. Future research could benefit from a more inclusive approach that encompasses a wider range of languages and access types to enhance the comprehensiveness of the findings.

### 3.2. Expert interview

The expert interviews played a crucial role in this study, serving to validate and refine the factors influencing the adoption of smart livestock technologies extracted from the scoping review. The choice of experts as the primary source of data was informed by several considerations, as follows.

System-Level Perspective: This study focuses on understanding the adoption of smart livestock technologies from a system-level perspective, particularly within the broader social, economic, and policy context of Japan. Experts in the cattle, swine, and poultry sectors possess a comprehensive view of the industry, including policy, economic trends, and technological advancements. Their insights are crucial for understanding the systemic factors that influence technology adoption within the specific context of Japan.

Expertise and Comprehensive Insight: Experts in the cattle, swine, and poultry sectors possess a deep understanding and comprehensive knowledge of the industry. Their expertise and familiarity with the challenges and future trends in the sector make them a valuable source of insights, especially given the complexity of the topic at hand.

Access to First-Hand Observations: Although farmers were not directly interviewed, the experts selected have had substantial interaction with farmers in their professional roles. This provides a unique opportunity to gain second-hand insights into farmers' experiences, behaviors, and attitudes toward smart livestock technologies.

Triangulation and Validation of Data: Expert interviews served as a mechanism for validating and refining the factors extracted from the scoping review. This triangulation process adds robustness to the research findings by cross-verifying data from different sources.

To ensure a comprehensive understanding, the interviews were structured in two parts. First, the factor cards derived from the literature review were used to verify existing knowledge and gather expert opinions on specific themes related to technology adoption. While there may be overlap between these themes and additional factors, the COM-B model was employed to capture Capability, Motivation, and Opportunity, providing a broader and more inclusive framework. This approach allowed us to identify a more complete set of influences on technology adoption among livestock farmers.

Despite the advantages, the use of expert interviews has limitations. Experts' perspectives, while invaluable, are not a direct substitute for the lived experiences of the farmers themselves. Their views may be shaped by professional biases or blind spots. Recognizing these limitations, the research findings were interpreted within the context of these potential biases.

Participants were recruited using convenient sampling and snowball sampling techniques, with a total of 10 experts participating. Table 2 shows the expertise of the interviewees.

Each participant was offered a monetary incentive of 3,000 Japanese yen per hour for their time and expertise. The interviews were conducted via Zoom, and the extracted factors from the scoping review were presented one at a time using Miro, a collaborative online platform. Prior to the interviews, factors influencing the adoption of smart livestock technologies were identified through a scoping review and presented on individual factor cards (a total of 20 cards). Each card represented one factor, as shown in Table 4. Participants were asked to read each factor card and provide any comments, revisions, or questions they had using digital sticky notes, as illustrated in Figure 1. Additionally, the interviewer asked follow-up questions based on the comments and discussion on the factor cards, and the participants' responses were recorded and transcribed for data analysis.

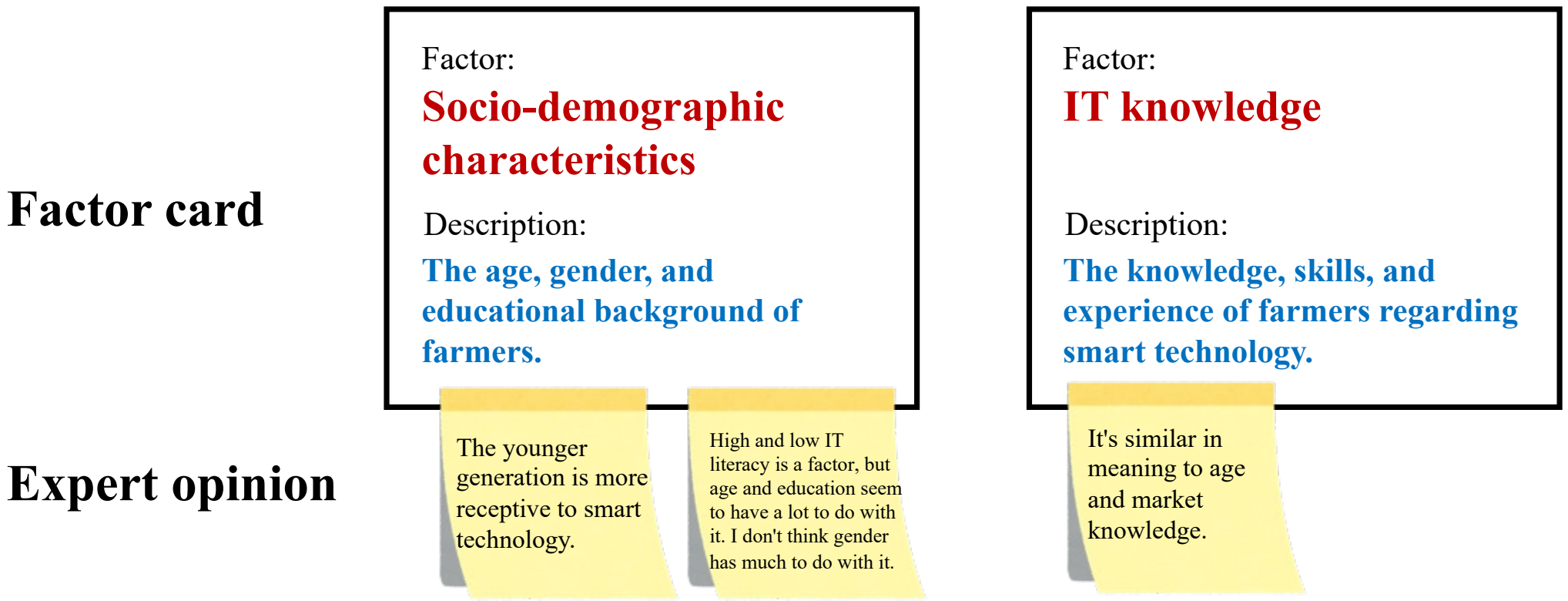

**Figure 1 Illustration of how extracted factors are presented in the expert interview and task.**

Once all factor cards had been discussed and the participants confirmed that no additional factors needed to be considered, the interviewer proceeded with semi-structured questions based on the COM-B model. These questions aimed to explore the capabilities, motivations, and opportunities that influence farmers' adoption of smart livestock technologies: (1) What skills, experiential barriers, or psychological hurdles do you think farmers face when adopting smart technologies? (capability); (2) What motivates farmers to adopt smart technologies? (motivation); (3) What external factors or triggers do you think influence farmers' decision to adopt smart technologies? (opportunity).

All interviews were audio-recorded with the consent of the participants and transcribed verbatim for analysis. Approval for conducting this study was granted by the Institutional Review Board at the Tokyo Institute of Technology (Approval No.: 2022196). Data collection continued until theoretical saturation was reached, i.e., no new themes or insights emerged from the interviews (Breckenridge and Jones, 2009).

Table 2 presents the professional backgrounds and roles, years of experience in the specific livestock species and production sectors, as well as the dates and times of the interviews.

One of the interviewees was an employee of a company that contributed research funds to this study. This has the potential to introduce bias in several ways, as follows.

Sampling Bias: The inclusion of a company employee in the pool of interviewees could potentially skew the data if their views are significantly different from those of the other interviewees. However, this risk is somewhat mitigated, as the employee represents only one-tenth of the sample.

Response Bias: The company employee might have consciously or unconsciously provided responses that favor the company's viewpoint or interests.

To minimize these potential biases, we took several precautions. Importantly, the interviewer had no prior relationship with the interviewee, an employee of the funding company. Moreover, the study's findings do not directly impact the company's marketing or operational activities, thus reducing the incentive for biased responses. Despite these precautions, readers should be aware of this potential conflict of interest when interpreting the study's findings.

**Table 2 List of research participants.**

| ID | Role; Specialization/job | Specialty livestock breeds | Years of experience in the livestock sector | Experience with farmers | Interview date and time |
|---|---|---|---|---|---|
| I01 | Researcher; Livestock Science, Livestock Management, Applied Animal Behavior | Dairy and beef cattle | 26 years | I01 has extensive experience since 2008 in smart livestock technology adoption, involving collaborative projects and demonstration experiments with farmers. I01 has also established cooperative relationships with Wagyu farmers, supporting the introduction of advanced systems. | May 2, 2022 10:15–11:45 |
| I02 | Sales; Livestock Management System | Swine, cattle, poultry | 11 years | I02 has extensive experience in promoting and selling smart agricultural technologies to farmers, emphasizing solutions that simplify management, enhance data analysis, and improve operational efficiency. | May 31, 2022 10:00–11:30 |
| I03 | Researcher; Animal behavior, animal welfare | Poultry | 15 years | I03 engages in smart agriculture technology through robotics and image analysis research, including developing a chicken care robot. I03 participates in exhibitions, interacting directly with farmers, and understand their needs and challenges, applying his/her research practically to deepen their involvement with the farming community. | July 1, 2022 9:00–10:30 |
| I04 | Researcher; Behavioral physiology, neurobehavioral science | Dairy and beef cattle | 29 years | I04 works closely with farmers to implement and refine agricultural technologies. Using farmer feedback, I04 develops technologies like stress measurement with ECG data. This collaboration ensures the technologies are practical and beneficial in real-world farming conditions. | July 4, 2022 9:00–10:30 |
| I05 | Researcher; Livestock Management, Farmer Welfare | Dairy cattle, poultry | Over 10 years | With over a decade of experience in dairy farming and a year each in layer hen and poultry technology, I05 has been supporting livestock farmers and has developed a keen interest in smart agriculture. From the perspective of farmers and animals, I05 focuses on technologies that can improve farmers' welfare and livestock management. | July 7, 2022 9:00–10:30 |
| I06 | Digital transformation strategy leader; Development and marketing of management technology | Swine | 4 years | Leveraging his/her experience in the feed industry, I06 has been dedicated to digitalizing the livestock sector. With a deep understanding of the challenges faced by farmers, I06 has provided solutions that contribute to labor savings on the ground, such as digitizing factories using tablets and developing non-contact pig weight measurement systems. Moreover, by quantitatively demonstrating the effects of implementing digital technologies, I06 has made the cost benefits clear for farmers and has been working towards the adoption of these solutions. | July 9, 2022 10:30–12:00 |
| I07 | Researcher; Livestock Management, Livestock Behavior, ICT Livestock Production | Beef cattle | Over 25 years | I07 has been actively working with local government officials and agricultural committees to build trust with farmers who practice grazing. To promote grazing, it is essential to gain the understanding of the local community, especially when it comes to addressing concerns about odors from livestock such as pigs. I07 has been persistent in explaining these issues to the community. Through these steady efforts, I07 appears to be deepening ties with farmers who engage in grazing. | July 11, 2022 9:00–10:30 |
| I08 | Technology developer; Development of management techniques | Dairy and beef cattle | 7 years | As the CTO of a smart livestock technology solutions company, I07 has been visiting farmers directly and engaging in face-to-face communication to gain a deep understanding of the challenges farmers face and their attitudes towards new technologies. Although sometimes confronted with harsh opinions, these experiences have provided valuable insights into farmers' perspectives and their evaluations of competitors' services. | July 12, 2022 17:00–18:00 |
| I09 | Incubator; Animal Science, Molecular Biology, Biotechnology | Swine | 21 years | With a background in livestock science and experience working on a pig farm, I09 has been involved in developing pig feed, launching branded pork businesses, collaborating with pig farmers, and supporting the introduction of smart livestock technology. I09 combines his/her hands-on experience in pig farming with knowledge of the latest technology, communicating it to farmers and facilitating collaboration. | August 26, 2022 9:00–10:30 |
| I10 | Cattle Specialist Veterinarian and incubator; Physiology, Breeding | Cattle | Over 10 years | I10 is a livestock expert specializing in cattle, with over a decade of experience in the field. I10 has focused on treating cattle diseases, managing feeding, and other veterinary tasks. Since joining an incubation company, I10 has been involved in supporting research at demonstration farms and promoting the social implementation of livestock-related technologies. | September 8, 2022 10:00–11:30 |

### 3.3. Modified-grounded theory approach

The present investigation made use of the M-GTA (Kinoshita, 2020) as an analytic lens through which to examine the collected interview data. It should be clarified that M-GTA is not intended to generate a formal theory, such as the TAM. Rather, it serves as a methodological tool intended for the development of domain-specific, or substantive, theories. These theories, while grounded in the data, carry the potential for limited generalization in contrast to the broad-based generalizations of abstract formal theories.

The TAM, for instance, is a formal theory that, while capable of providing insights into smart technology adoption within the livestock industry, is often criticized by practitioners because of its highly abstract nature (Lee et al., 2003). Its lack of specific guidelines for designing effective systems or choosing among competing systems often leaves practitioners feeling dismissed or underrepresented, despite the simplicity of the model, which might appeal to academic researchers (Ohashi et al., 2021).

On the other hand, by its constructivist design, M-GTA fosters the development of theories firmly grounded in the context-specific realities of the domain under investigation. Consequently, it facilitates a nuanced, contextually relevant, and practically useful understanding of the smart technology adoption process in the livestock sector. As a methodology, it was thus deemed suitable for this study, given its ability to provide a level of detail and specificity that can be of high utility to diverse stakeholders, such as policymakers, technology developers, and other practitioners.

In the application of M-GTA, a series of steps were followed to ensure that the substantive theory emerging from the data remained robust, trustworthy, and demonstrably relevant to the specifics of the domain under investigation. These steps comprised the definition of analytical themes, construction of an analysis worksheet, generation of concepts, comparison of counterexamples, generation of categories, and creation of a result diagram and storyline. To further substantiate the study's credibility and trustworthiness, peer debriefing was integrated into the methodology (Janesick, 2015). This process involved the critical evaluation and discussion of the initial grounded theory among all authors, thereby ensuring the robustness of the theory and confirming its alignment with the data. Detailed steps are as follows.

1. Analytical Theme Definition: Analytical themes were established, referring to research questions elucidated through analysis and identifying specific processes. This study defined the analysis theme as "the process of smart technology adoption by Japanese livestock farmers."
2. Analysis Worksheet Construction: An analysis worksheet was generated (Table 3), encompassing concept names, definitions, variations (specific utterance examples), and theoretical notes. Here, a concept signifies the abstract notion of the phenomenon, structure, or relationship under examination. A definition clarifies a concept's meaning and scope. A variation represents a tangible speech example, transcribed from an interview transcription. A theoretical memo explicates and theoretically interprets insights, findings, and relationships acquired during data analysis. The variation column has been labeled "(Extracted-factors_ID)" or "(ID_No.)" to correspond to the interview transcript. The former identifies the interviewee ID for extracted factors in the scoping review, and the latter is used in the COM-B-based interview transcript to track whose statements are in the variation by entering the interviewee ID and the column number of the verbatim transcript.
3. Concept Generation: The first author examined transcriptions based on the analysis theme, identified utterances seemingly generating the concept, and populated the variation columns. Subsequently, the sentence's meaning was contemplated, and the definition column was filled with a concise statement. A word encapsulating the concept's content was conceived, designated as a concept, and entered into the concept column. This interpretive task sequence generated diverse ideas and questions, documented in a theoretical memo. The data were then examined for specific utterance examples aligning with the generated definitions. Any discovered examples were noted in the variation column similarly. Once sufficient examples emerged within the variation column, the concept was deemed valid. Typically, multiple utterance examples were required to validate a concept. However, in exceptional cases, a single utterance was considered sufficient if it was particularly insightful or represented a critical aspect of the phenomenon under study. For example, an utterance that uniquely captures a key challenge or breakthrough in smart technology adoption could be deemed significant enough to stand alone as a concept.

4. Counterexample Comparison: Concrete counterexamples to the generated concept definitions were sought, enabling the identification of the maximal possible range of phenomena while precluding the generation of theory exceptions.
5. Category Generation: Relationships between generated concepts and other concepts were individually examined, producing higher-level concepts known as subcategories, and subsequently, their higher-level counterparts, categories. These relationships were summarized in a relationship diagram.
6. Result Diagram and Storyline: Relationships among generated categories, subcategories, and concepts were individually investigated, and these relationships were documented as a result diagram. Additionally, a succinct result summary was recorded as a storyline.
7. Peer Debriefing: To further enhance the credibility and trustworthiness of our study, we engaged in the process of peer debriefing (Janesick, 2015). After generating the initial grounded theory, we discussed the theory with the other co-authors (i.e., the second, third, and last authors). In preparation for these discussions, the first author shared the list of categories, subcategories, the relationship diagram, and the storyline with the co-authors. These data were shared in advance, and we then held detailed discussions either face-to-face or through Zoom video conferencing. The generated theory and storyline were presented and critically evaluated during these sessions. Each author critically reviewed the findings and proposed revisions where necessary. This collaborative discussion and revision process enhanced the robustness of the theory and ensured that the theory adequately reflected the data. Disagreements were resolved through consensus, and the theory was refined based on these discussions.

In this study, the M-GTA analysis commenced after the completion of all ten interviews. Each interview transcript was analyzed sequentially for concept generation, with continuous comparison performed throughout the process. This iterative approach allowed for the integration of new insights and the refinement of concepts as each interview was analyzed. Theoretical saturation was considered achieved when no new concepts emerged from the final interview transcript, and all concepts, subcategories, and categories were interconnected. In this research, theoretical saturation was attained as no new concepts were identified in the tenth interview. Had new concepts emerged at that stage, additional interviews would have been conducted to ensure comprehensive analysis and achieve theoretical saturation.

Applying this approach to the interview data, we were able to derive a comprehensive and robust process model that captures the factors affecting the adoption of smart livestock technologies.

**Table 3 A template of the analytical worksheet used in this study.**

| Concept name | Definition | Theoretical memo | Variations |
|---|---|---|---|
| | | | Interviewee's utterances for extracted concepts (EXTRACTED FACTORS_ID)<br>Interviewee's utterance for COM-B (ID_No.) |

### 3.4. Researcher Characteristics and Reflexivity

The research team consisted of four researchers with diverse backgrounds and expertise. The first author has substantial experience conducting scoping reviews and applying human-centered design across various fields, including livestock farming. The second author, while not possessing extensive knowledge in livestock-related areas, holds a Ph.D. in Applied Linguistics and brings expertise in discourse data analysis. The third and fourth authors are deeply engaged in developing smart technology for swine production and have extensive industry contracts.

In line with the constructivist worldview, we acknowledge that our diverse backgrounds and experiences influenced each stage of our research process, from the framing of questions to the interpretation of findings. To mitigate potential biases and ensure a balanced interpretation of the data, we utilized the M-GTA, characterized by its constant comparison nature. This iterative process of data collection and analysis allowed us to continuously reflect on our perspectives and reconsider our assumptions. It served as a tool to promote our reflections on potential biases and maintain our focus on the viewpoints of the research participants. Additionally, peer debriefing sessions served as a platform for us to critically review our findings, challenge each other's interpretations, and make necessary revisions.

# 4. Results

## 4.1. Factors influencing the adoption of smart technology in livestock production: A scoping review

Figure 2 shows the flowchart of the scoping review process according to the PRISMA-ScR guidelines. A comprehensive search of Web of Science All Databases yielded 291 articles, which underwent a rigorous screening process based on predefined inclusion and exclusion criteria. Specifically, only original, English-language, open-access journal articles were selected, resulting in 85 articles. The first author reviewed the titles and abstracts of these papers to identify those that met the criteria listed in Table 1, resulting in a total of 19 papers for full-text screening. Of the 19 papers, 10 were excluded, including 1 review paper, 1 irrelevant to livestock farming, and 8 not related to smart livestock technology. Finally, 9 papers were included in the analysis. The first author performed open coding on the selected papers to identify factors influencing the adoption of smart technologies in livestock production, resulting in 46 concepts. Further analysis using second-order coding yielded a total of 20 factors, which are summarized and presented in Table 4 with their corresponding descriptions and source information.

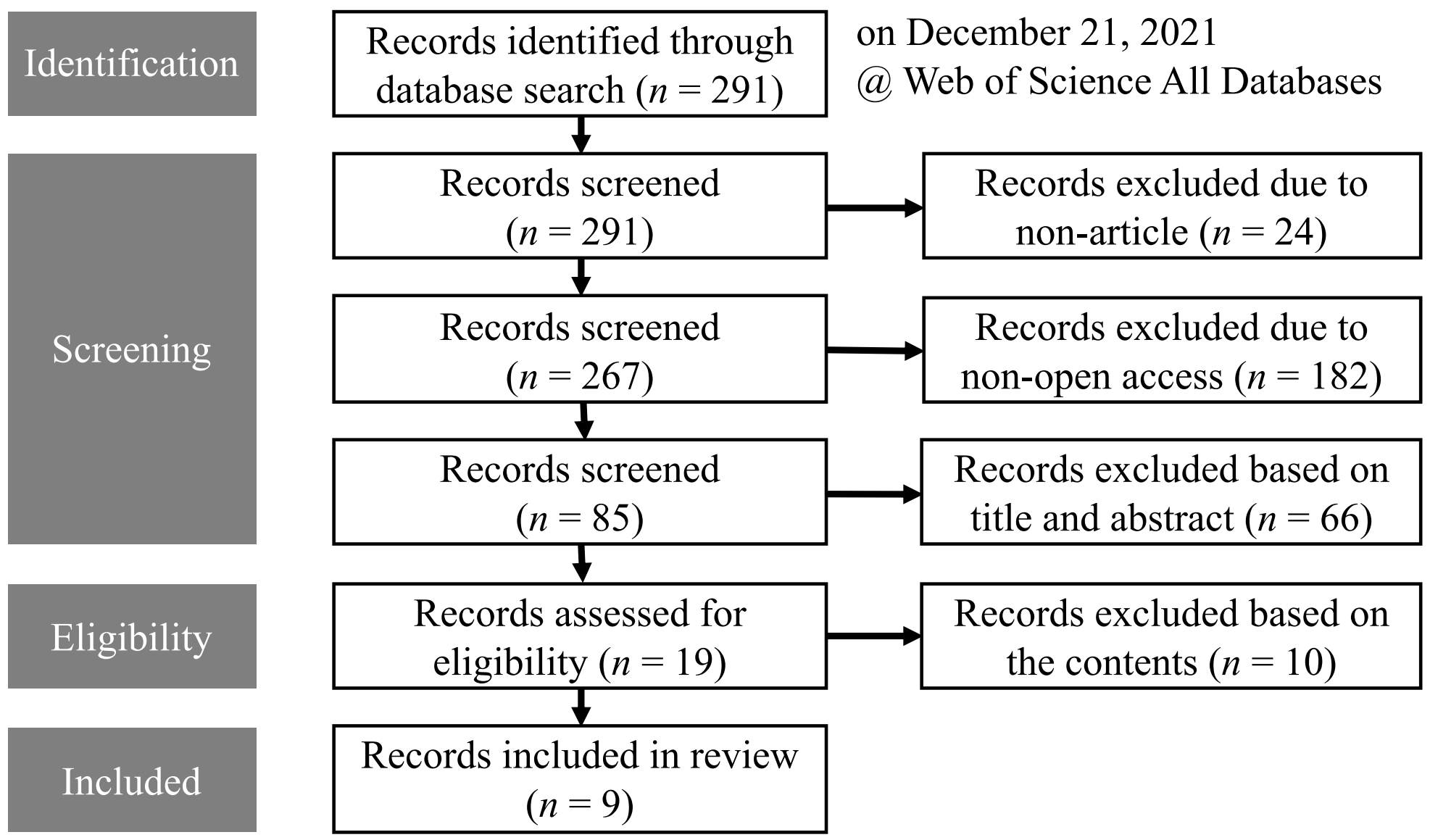

**Figure 2 PRISMA-ScR Flowchart.**

**Table 4 Extracted factors associated with smart technology adoption by livestock farmers from the scoping review.**

| Extracted factors | Description | References |
|---|---|---|
| Socio-demographic characteristics | The age, gender, and educational background of farmers. | (Filippini et al., 2020; Groher et al., 2020; Liu et al., 2019; Michels et al., 2019) |
| Agri-environmental scheme membership | The status of grant acquisition for farmers engaged in agriculture that takes into consideration biodiversity, landscape, and environmental conservation. | (Liu et al., 2019) |
| IT knowledge | The knowledge, skills, and experience of farmers regarding smart technology. | (Kaler and Ruston, 2019; Khanal et al., 2010; Lima et al., 2018; Michels et al., 2019) |
| Market knowledge | The knowledge, skills, and experience related to the market for developing animal husbandry as a business. | (Kebebe, 2019) |
| Entrepreneurship | The intention to expand the farm size and the inclination to develop and experiment with new products and services without missing out on business opportunities. | (Kebebe, 2019; Lima et al., 2018) |
| Perceived value | The degree to which farmers perceive value in smart technology, such as compatibility with traditional work styles and the potential for high returns from technology adoption. | (Kaler and Ruston, 2019; Todeschini et al., 2020) |
| Perceived usefulness | The degree to which farmers perceive smart technology as useful in improving productivity, for example, by increasing efficiency or reducing labor requirements. | (Kaler and Ruston, 2019; Lima et al., 2018; Michels et al., 2019) |
| Perceived ease of use | The degree to which farmers perceive smart technology as user-friendly and easy to use. | (Lima et al., 2018; Michels et al., 2019) |
| External pressure | External pressure from organizations to adopt smart technology. | (Lima et al., 2018) |
| Belief about importance of human intervention | Farmers' belief that human intervention is necessary for good animal husbandry practices. | (Kaler and Ruston, 2019) |
| Perceived room for investment | Financial capacity to invest in the adoption of new technology. | (Kebebe, 2019; Liu et al., 2019) |
| Location | The region and topography of the farm (e.g., hilly terrain, mountainous terrain, or basin). | (Groher et al., 2020) |
| Management type | Livestock management methods (e.g., tethered, free-range, or grazing) and farming types (e.g., livestock, mixed grain/livestock). | (Groher et al., 2020) |
| Farm size | The size of the farm and the number of animals raised. | (Groher et al., 2020; Khanal et al., 2010) |
| Production yield and income | Agricultural income or production volume (e.g., milk production). | (Liu et al., 2019; Michels et al., 2019) |
| Livestock breed | Dairy cows, beef cattle, pigs, laying hens, broilers, etc. | (Groher et al., 2020) |
| Production degree centrality | The extent to which the farm plays a central role in regional production, for example by providing or lending equipment to other farmers. | (Filippini et al., 2020) |
| Access to technological information | Ease of access to new technology information (e.g., information disclosure, relationships with research institutions). | (Kebebe, 2019; Liu et al., 2019; Todeschini et al., 2020) |
| Access to financial support/loan | The level of financial support from the government and other entities, and the ease of obtaining credit for new investments. | (Kebebe, 2019) |
| Support institutes/programs | Non-financial support and programs provided by government or administration. | (Kebebe, 2019) |

### 4.2. Livestock farmers' smart technology adoption process

In this section, we present the results of the M-GTA analysis of the interview transcripts with the experts. The analysis led to the generation of 84 initial concepts, detailed in the analytical worksheet in Table A1. These concepts were systematically compared and grouped into 25 subcategories, which are presented in Table 5. These subcategories represent a higher level of abstraction, summarizing multiple related concepts. Further abstraction and integration of both subcategories and any remaining ungrouped concepts resulted in the formation of 10 overarching categories, as shown in Table 6. This hierarchical organization from concepts to subcategories and finally to categories provides a structured understanding of the factors influencing smart technology adoption among livestock farmers.

**Table 5 Subcategories generated by concepts comparison.**

| Subcategory | Description | Encapsulated concepts |
|---|---|---|
| Advanced management mindset | Livestock farmers have a sense of awareness of the issues and are emphasizing the importance of production efficiency, profits, and consideration for the environment from a corporate management perspective. They are adopting a proactive and motivated approach toward the introduction of smart technology, with a recognition of the economic benefits it brings. | Awareness of the issues, Corporate management perspective, Recognition of economic benefits |
| Animal welfare transition | Livestock farmers undergo a process of changing their values, beliefs, husbandry practices, and production systems in order to recognize the importance of animal welfare, and work toward environmental conservation and productivity improvement. This process involves the adoption of smart technologies, which have different relationships with each type of livestock, and are influenced by changes in husbandry practices and market trends. Pressure for animal welfare and government support also play important roles. | Paradigm shift to animal welfare, Discrepancy between animal welfare compliance and smart technology, Transitional period in animal care, Pressure to comply with animal welfare standards |
| Beliefs in livestock breeding methods | The beliefs and values related to environmental considerations, animal welfare, and breeding methods held by livestock farmers, which influence the adoption and utilization of smart technology. | Low interest in environmental considerations, Interest in living things, Farmers' beliefs about breeding methods |
| Collaboration challenges among stakeholders | The difference in knowledge and experience between government support measures, agricultural cooperative officials, and farmers can affect collaboration, leading to challenges in providing sufficient and effective support and advice. | Lack of response to administrative support for subscription services, Knowledge gap between farmers and agricultural cooperative officials |
| Complexity | The phenomenon of complex judgment elements involved when adopting smart technology. These elements affect the appropriate technology selection and dissemination for livestock farmers. | Difficulty in selecting competitive products, Tradeoffs in technology adoption |
| Conservatism | Livestock farmers are reluctant to adopt new smart technologies, characterized by attitudes and beliefs that emphasize the importance of face-to-face communication, traditional agricultural values, and aversion to new things. | Importance of face-to-face communication, Traditional view of agriculture, Aversion to new things |
| Desire for growth | Livestock farmers have a strong motivation to adopt new technologies, adapt to market trends, and pursue innovative business models and expansion. | Promoting the entry of young people, Challenges for new business and business expansion, Response to market trends |
| Digital literacy | Necessary skills and knowledge to adapt to new digital technologies, such as the ability and attitude to utilize smart technologies. Digital literacy is an important factor in the smart technology adoption process for livestock farmers, and it varies depending on age and educational level. People with high digital literacy tend to actively embrace new technologies, which may facilitate the introduction and use of such technologies. | Self-efficacy of technology use, Familiarity with smart technology, Information-gathering ability |
| Dynamics of family-owned businesses | A concept that considers how elements such as the tension between tradition and innovation, the role of decision-makers, and family understanding within a family-run business may influence the adoption process of smart technology. This category provides a framework for understanding how the characteristics and power dynamics of family-run businesses are involved in the success or failure of the introduction of smart technology. | Craftsman-like father block, Understanding of family, Conflict between family business tradition and innovation |
| Ease of implementation | Elements related to the perceived added value, such as ease of installation and operation, and suitability for business flow, that are important to consider when introducing smart technology. | Ease of installation, Fit with on-site routine |
| Ease of use of technology | In the process of technology adoption, not only the performance of the technology but also the ease of operation, the readability of the screen, and the reduction of inconvenience are taken into consideration. This subcategory includes challenges for promoting technology diffusion among different socio-demographic groups (e.g., women, elderly people, etc.). | Technological adoption barriers for women, Ease of use of technology |
| Economic leeway | The concept that indicates the degree to which factors related to investment capacity, such as economic constraints, risk-aversion attitudes, business scale, income levels, etc., influence the technology adoption process when introducing smart technologies into the livestock industry. | Lack of investment capacity, Investment capability |
| Expectations for improved productivity and labor-saving measures | Livestock farmers expect to increase productivity and reduce labor by introducing smart technology, which can formalize implicit knowledge, improve economic efficiency, increase work efficiency, and increase profits. | Expectations for formalizing tacit knowledge, Expectations for productivity improvement, Expectations for labor saving |
| Facility renewal and generational change | The process of creating opportunities for the introduction of smart technology arises when there is a deterioration or expansion of facilities or a generational change due to business succession. In particular, younger generations tend to utilize IT and new technologies, which promotes the adoption of smart technology in business succession. | Facility renewal opportunities, Succession of business to younger generations |
| Formation of attitudes through technical experiences | The process of forming attitudes toward the adoption and adaptation of new technologies, which are influenced by the positive or negative impact gained from the introduction and use of smart technology, as well as the knowledge and experience obtained from them. | Shift in values due to technological adoption, Positive experience from existing systems, Negative buying experience |
| Impact of agricultural policies | The impact of government policies, initiatives, and financial support on the adoption of smart technologies by livestock farmers. This includes government efforts to promote technology dissemination and provide subsidies and grants to alleviate the adoption willingness and economic burden of farmers. However, effective communication and | Administrative impact, Impact of financial support |
| | collaboration between farmers and the government are important, and financial support alone may not always be sufficient. The differences in subsidies based on livestock type and accessibility to information are also considered as factors affecting the adoption of technology. | |
| Infrastructure constraints | Phenomena that create challenges and constraints related to land use and information infrastructure development. These include land conditions, underdeveloped information infrastructure, quarantine and environmental issues. These constraints affect the location conditions and business processes of livestock farmers and can negatively impact the adoption and operation of smart technologies. | Constraints on land use, Inadequate information infrastructure |
| Limited access to information | Livestock farmers face difficulties in accessing and understanding information on the adoption and application of smart technology. | Information disconnection, Restrictions on information exchange, Lack of transparency in individual management information, Information gathering from industry publications |
| Livestock management systems | Factors that affect the adoption process and adaptability of smart technology in various breeding methods and environments employed by livestock farmers. The farm size, management style, and animal species also influence the breeding methods. | Small-scale and niche breeding methods, Breeding methods suitable for precision management |
| Maintaining the status quo mindset | Livestock farmers are generally hesitant to adopt smart technologies, and tend to stick to current breeding systems and management styles, resulting in a tendency to be dependent on current market trends. | Tendency to maintain the status quo, Inertia, Dependence on market prices |
| Observability of the effect | The degree to which the effects and benefits can be observed through implementation. Observability is an important factor when considering adoption or continuation, and the immediacy and concreteness of the effects vary depending on the livestock species. | Reflection of efforts, Time lag to realize value, Cost-effectiveness of technology adoption for different livestock species |
| Structural challenges in the livestock industry | The livestock industry is facing various problems and challenges, such as a situation where the efforts of livestock farmers do not correlate with their profits, a lack of appropriate support for introducing technology due to a shortage of manufacturers, low IT literacy among distributors, etc. These issues may reduce the willingness of farmers to adopt new technologies and adapt to changes, which could potentially impact the spread of smart technologies in the industry. | Income not proportional to effort, Lack of sales and development manufacturers, Decline of the industry as a whole |
| Technology implementation support | A series of support activities provided to livestock farmers for effective adoption of smart technologies, aiming to solve problems and improve productivity on-site. This includes technology dissemination, training, guidance, implementation support, and communication with livestock-related companies. The main supporters of these activities are the government, manufacturers, feed companies, and smart technology vendors. | Support organizations and programs, Technology education and support |
| Time and psychological leeway | The degree of psychological leeway that affects the adoption of technology because of the workload or personal circumstances, while also being a factor that creates a positive attitude toward technology adoption and encourages experimental approaches, as it allows for more time flexibility | Psychological allowance, Lack of psychological allowance, Lack of flexibility due to workload |
| Word-of-mouth effect | This refers to the impact that information sharing and evaluation from influential producers, veterinarians, feed companies, and other farmers have on decision-making in the adoption process of smart technology. This can result in the promotion or inhibition of the spread of new technologies. | Influence from influential producers, Influence of word-of-mouth on decision-making |

**Table 6 Categories generated by subcategories and concepts comparison.**

| Category | Description | Encapsulated subcategories and concepts |
|---|---|---|
| Agricultural policy | This category encompasses policy factors, such as the government's guidelines and initiatives that affect the adoption process of smart technologies in livestock farming, financial support, and support systems. This can promote or hinder the adoption of smart technologies by livestock farmers. | Impact of agricultural policies, Collaboration challenges among stakeholders |
| Assessment and interpretation of technology | A comprehensive understanding of the expectations and evaluation criteria for adopting smart technology by livestock farmers, as well as the advantages and challenges of technology from various perspectives, and the process of interpreting and applying it to address them. | Expectations for improved productivity and labor-saving measures, Ease of use of technology, Ease of implementation, <Expectations for ability expansion>, <Expectations for problem solving>, <Durability of technology>, <Expectations for margin creation>, <Product reliability> |
| Availability of technology | The introduction and application of smart technologies are factors that indicate the extent and conditions in which they are possible, and these are influenced by factors such as livestock breeds, rearing environment, products, rearing methods, information infrastructure, and land use restrictions. | Infrastructure constraints, Livestock management systems, <Availability of smart technology> |
| Farm management policy | The business values and strategies based on multiple factors to decide the adoption of smart technology in livestock farming, including internal factors such as the age, experience, and education level of the business owner, and external factors such as market prices and competitive environment. | Dynamics of family-owned businesses, Conservatism, Advanced management mindset, Maintaining the status quo mindset |
| Livestock industry trends | In the livestock industry, there are complex factors such as the importance of animal welfare, market trends, technological innovation, and addressing consumer needs that are involved in phenomena and trends. These factors not only affect the smart technology adoption process of livestock farmers but also contribute to the overall sustainability and development of the industry. | Animal welfare transition, Structural challenges in the livestock industry, <Attention of the world> |
| Motivation | The concept that drives livestock farmers to adopt smart technology, aiming to achieve goals such as efficiency and competitiveness, is motivated by factors such as promoting young people's entry, challenging new or expanded business, responding to market trends, and hedging risks. | Desire for growth, <Risk hedge> |
| Opportunities for introduction | In the livestock industry, there are situations and timing where it is easier to introduce smart technology on farms. Factors such as facility upgrades or expansions, business succession, and psychological and economic leeway are involved, and combining these factors can promote the adoption of smart technology, leading to improved management efficiency and productivity. | Time and psychological leeway, Economic leeway, Facility renewal and generational change |
| Social relations | Social relationships have an impact on the decision-making of livestock farmers. Social relationships include production areas, production networks, veterinarians, feed companies, etc., and word-of-mouth, advice, evaluations, and the like from these relationships can influence the decision-making of livestock farmers. | Word-of-mouth effect, <Sense of competition with neighboring farmers> |
| Technology implementation | The process of considering, evaluating, and applying smart technology by livestock farmers, which includes learning from past experiences or existing systems, making final decisions on adoption or abandonment, and forming new attitudes based on the results of implementation. | Formation of attitudes through technical experiences, <Introduction of smart technology>, <Withdrawal of introduction> |
| Values of farmers | The unique attitudes and beliefs held by livestock farmers include elements such as environmental consideration, animal welfare, breeding methods, and social norms. The values of farmers influence the adoption and implementation process of smart technologies, and form the basis for promoting improvements in agricultural practices, sustainability, and efficiency. | Beliefs in livestock breeding methods, <Adherence to social norms>, <New entrants from other industries> |

Note: Encapsulated subcategories are displayed as is, and included concepts are displayed with < > brackets.

Based on the analysis worksheet, the relationships among each concept, subcategory, and category were examined and summarized in a process diagram (Figure 3). For simplicity, the categories, sub-categories, and uncategorized concepts were used in the process diagram, and categorized concepts were omitted. The following storyline is based on the process diagram:

The adoption of smart technology in livestock farming is heavily influenced by individual farmers' values, which interact with the farm management policy. Both farmers' values and farm management policies are shaped by social relations with other farmers, veterinarians, and feed companies, as well as by livestock industry trends and agricultural policies. In family-run farms, there are dynamics where individuals with an advanced management mindset may clash with those holding more conservative views. Farms with a conservative approach tend to maintain the status quo, while those with an advanced management mindset are more motivated to adopt smart technology. This motivation is influenced by social relations and the decision-makers' digital literacy levels.

Motivation and opportunities for technology introduction are closely linked. Due to the significant investment required for smart livestock technology, motivation can wane without clear opportunities for introduction. Agricultural policies, such as subsidies, play a crucial role in creating these opportunities. Furthermore, as agricultural policy varies depending on the livestock management systems, especially the livestock breed, the impact on the opportunities for introduction differs. The fit between the technology and the farm's management system, as well as existing infrastructure constraints, also affects technology availability and adoption opportunities.

When conditions for technology introduction are met, the farm transitions to the phase of assessing and interpreting the technology. Social relations, such as word-of-mouth from trusted farmers, greatly influence this phase. Positive social influences can reduce perceived risks, facilitating the transition. During the assessment phase, evaluations of the technology's usability and expected benefits are conducted. This phase is influenced by the farmer's digital literacy and the observability of the technology's effects. Effective technology implementation support can reduce the perceived complexity of the technology, leading to more favorable assessments.

Ultimately, the assessment phase leads to decisions about technology adoption or abandonment. Experience with the introduced technology shapes new attitudes toward it. High observability of positive effects fosters a favorable attitude, reinforcing ongoing adoption or further technology integration. This evolving attitude can, in turn, influence farmers' values.

Farmers' values and farm management policies, collectively represented by their social relations, interact with livestock industry trends. These trends influence the motivation for adopting farm technology. Moreover, livestock industry trends significantly impact the adoption process of smart technology through their interaction with agricultural policy.

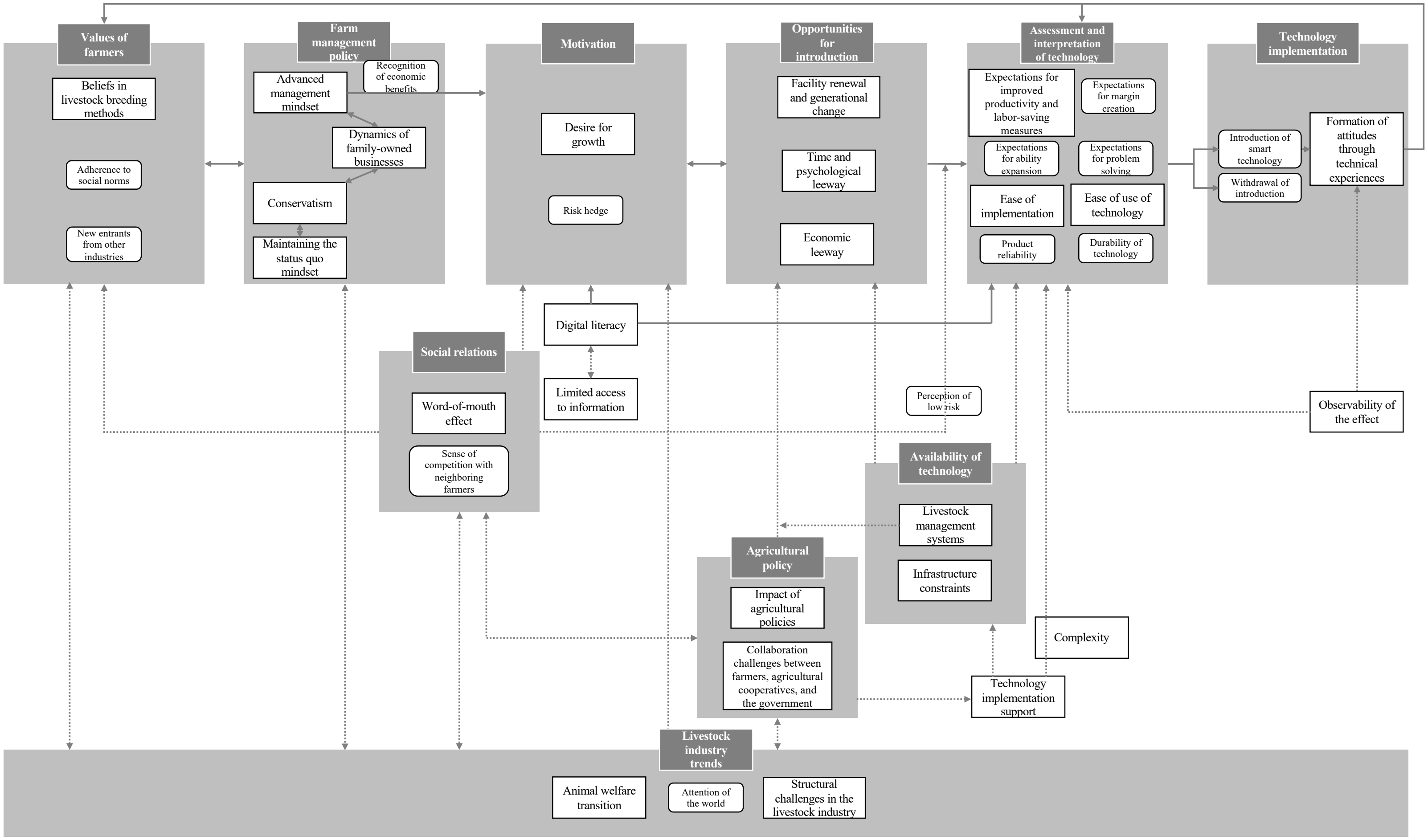

**Figure 3 Process diagram of smart technology adoption by livestock farmers. Solid lines represent causality, relationships, and influences within the farm context. Dashed lines denote external causality, relationships, and influences, either external to the farm or from the external environment to the farm. Grey boxes represent categories, rectangular boxes denote subcategories, and rounded rectangles symbolize concepts.**

# 5. Discussion

## 5.1. Theoretical implications

### 5.1.1. Multi-level perspective on-farm decision-making on technology adoption

The terminology shift from "technology adoption" to "farm decision-making" in our study is an important distinction that requires further clarification. While the terms are often used interchangeably, they denote subtly different aspects of the same process. "Technology adoption" tends to focus on the outcome—the decision to adopt or not adopt a given technology. "Farm decision-making," on the other hand, encompasses a broader range of considerations, of which technology adoption is a significant part.

Our process diagram provides a structural understanding of the decision-making process surrounding the adoption of smart technology in farming. We suggest the categorization of the process into three interconnected levels: farm level, socio-technical level, and landscape level, each of which dynamically interacts in shaping the decision-making process.

Farm-level decision-making is deeply rooted in individual farmer values and farm management policies (Burton et al., 2008). Yet, these decisions are far from autonomous, as our findings illustrate their interdependence with larger socio-technical factors and industry trends.

The socio-technical level includes elements like social relations, availability of technology, and agricultural policy, which exert a substantial influence on farm-level decisions. Our study underscores the role of the word-of-mouth effect from trusted farmers and community members, which substantially impacts a farmer's technology assessment and interpretation. This observation aligns with Latour's actor–network theory, suggesting that social relations play a crucial role in shaping individual behaviors (Latour, 2007). For instance, our findings indicated that the influence of trusted farmers can foster a sense of low-risk perception, promoting the transition to the next phase of technology adoption. In Japan, the impact of traditional agricultural protection policies and the presence of agricultural cooperatives have resulted in relatively slow farm consolidation compared to Western countries, maintaining a higher proportion of small-scale farms. Ichida et al. (2009) highlight the importance of analyzing smaller community units in Japan, as opposed to the large-scale regional studies typical in Western countries. In Japan, these smaller community units significantly impact the psychology and behavior of residents due to their strong social cohesion. This suggests that strong social ties and communal relationships within small communities create complex dynamics unique to the Japanese context. This suggests that strong social ties and communal relationships within small communities create complex dynamics unique to the Japanese context. For example, our study identified the concept of the "importance of face-to-face communication" (Table A1), which falls under the subcategory of conservatism (Table 5). This concept reflects a preference for prioritizing the attitudes and behaviors of technology vendors over the technical performance of the technology itself. Such a tendency is more prevalent in countries like Japan, where small-scale farmers form close-knit communities, in contrast to Western countries, where large-scale, corporate-managed farms dominate. Consequently, the emphasis on "importance of face-to-face communication" varies by country, shaped by national agricultural policies that influence the social relationships surrounding farms.

The landscape level encompasses broader industry-wide factors, such as societal concerns reflected in trends like animal welfare transitions. These trends have an indirect but potent impact on the farm-level decisions. For instance, as societal concerns for animal welfare grow, these may influence agricultural policies promoting welfare-friendly technologies. Such policies can impact farm management policies, thereby highlighting the complex interconnectedness of these levels (Geels, 2002; Ingram, 2008). In our study, several experts highlighted the impact of animal welfare, indicating that Japan is in a transitional phase concerning animal welfare. Specifically, the awareness rate of animal welfare among Japanese consumers is approximately 10-20% (Washio, 2023), suggesting limited recognition of the concept. Additionally, international pressures within the context of ESG investments have prompted responses from both the government and major corporations. In 2023, MAFF introduced new guidelines on animal welfare, reaffirming Japan's commitment to meeting international standards, such as those outlined in the OIE codes, in light of global trends like the expansion of livestock product exports and the increasing significance of the United Nations' SDGs (MAFF, 2023). Furthermore, major Japanese food processing companies have initiated efforts to address animal welfare, with one company aiming to eliminate gestation crates for pigs by 2030 (NH Foods Ltd., 2023). Compared to Europe, where animal welfare practices are well established, Japan's transitional phase suggests that animal welfare trends and related policies will have a substantial impact on the adoption of smart technologies in the future. As animal welfare becomes more widely recognized outside of Europe, its influence is expected to extend to the context of smart technologies, particularly in countries where animal welfare has not yet become a firmly established social norm or regulatory framework.

The aforementioned discussions align with the Triggering Change Model, which elucidates that innovation diffusion hinges upon the unique innovation environment of farmers (Sutherland and Labarthe, 2022). However, our study extends this perspective by considering livestock industry trends, which may shape the innovation environment of farmers. This expanded view echoes the multi-level perspective that recognizes the interaction between multiple levels in shaping innovation trajectories (Geels, 2002; Sutherland et al., 2015).

By explicitly recognizing and examining the broader “farm decision-making” process, our study provides a more holistic understanding of the adoption of smart technology in the livestock sector. This approach acknowledges that the decision to adopt a specific technology is influenced by myriad factors beyond the technology itself, including individual values, social relations, and broader industry trends.

In conclusion, our multi-level perspective contributes to a more comprehensive understanding of the decision-making process in livestock farming. It accentuates the complexities involved in adopting smart technology within this sector, emphasizing that these decisions result from an intricate interplay between individual, social, and institutional factors. This perspective could pave the way for future research on technology adoption in farming and, more practically, can inform policy-making, enabling the creation of supportive policies that consider the complex interplay of these factors. Additionally, it could guide farm advisory regimes in tailoring their services to better meet the needs and realities of farmers navigating this multi-level decision-making process.

#### 5.1.2. Sequential and iterative nature of technology adoption process

Our study provides a nuanced understanding of how farmers navigate the introduction of smart technology, highlighting the sequential and iterative nature of the technology adoption process. In contrast to the structure of the COM-B model (Michie et al., 2011), which suggests that capability, opportunity, and motivation operate in parallel to drive behavior, our findings emphasize a more sequential process. The COM-B model postulates that capability, opportunity, and motivation changes simultaneously influence behavior. However, our findings indicate that these factors might unfold in a sequential manner. For instance, we observed that a farmer might have the motivation to adopt new technology, but without the available opportunity, the farmer may not transition to the phase of considering and adopting the technology. This sequential nature is exemplified by our observed dynamics, where the motivation to adopt smart technology tends to weaken without the opportunities for introduction, which, in turn, are influenced by elements such as agricultural policy and livestock management systems.

This sequential aspect of technology adoption could be attributed to the unique characteristics of the livestock industry, which often requires substantial investment in land, infrastructure, facilities, and technologies. Such substantial investments may lead to technological lock-in (Liebowitz and Margolis, 1995; Sutherland and Labarthe, 2022), limiting opportunities for technology introduction based on livestock management systems and breeds. We observed this in instances where the nature of the technology being considered did not fit the livestock management systems, reducing the opportunities for the introduction.

Our findings align with the observation of Sutherland and Labarthe (2022) of a stepwise process in farmers' adoption of innovations in European agriculture. They implied that farmers first perceive a need or opportunity, then consider possible solutions and actively assess them, finally deciding whether to adopt an innovation based on its perceived advantages and their ability to implement it. This aligns with our study, where the transition from the opportunity to technology assessment was greatly influenced by social relations and the word-of-mouth effect from trusted farmers.

However, the adoption process we identified is not linear but displays an iterative nature, particularly in the formation of attitudes through technical experiences. This iterative aspect suggests a departure from traditional theories such as the TAM (Davis, 1985), which views technology adoption as a relatively linear process. In our study, farmers formed attitudes toward smart technology through their experiences of using it, and these attitudes evolved over time and influenced future adoption decisions. This echoes the findings of Sutherland and Labarthe (2022), who noted that the innovation implementation and consolidation process could lead to a reconsideration of options because of arising issues.

In conclusion, our study contributes to the literature on technology adoption in livestock farming by providing a nuanced understanding of the sequential and iterative nature of this process. It suggests that existing technology adoption models could be enriched by incorporating a more explicitly sequential and iterative perspective, where motivations, opportunities, and behaviors interact and evolve over time. This approach acknowledges the complex interplay of factors that shape the technology adoption process in the livestock industry, taking into account farmers' unique characteristics and experiences. By shedding light on the dynamic aspects of technology adoption, our findings can better inform policymakers, technology developers, and other stakeholders in promoting the adoption of smart livestock technologies.

### 5.1.3. Dynamics of on-farm decision-making on technology adoption

Our research illuminates the unique “dynamics of family-owned businesses” within the context of smart technology adoption in livestock farming. This empirical exploration contributes to family business theories, providing valuable insights into the interplay between an “advanced management mindset” and “conservatism,” and its consequential impact on technology adoption decisions within family-managed farms.

Wilson (2008) discusses the importance of orientation in decision-making, introducing the concept of decision-making corridors. This is explained as a divergence in the opportunities for decision-making afforded by productivism and non-productivism orientations. In our study, this difference in orientation can be interpreted as creating a variance in the presence or absence of opportunity via motivations.

Our findings underscore that the clash between the “advanced management mindset” and “conservatism” can significantly affect the technology adoption process, potentially inciting conflicts, causing delays, or necessitating compromises. Family-managed farms with a well-developed “advanced management mindset” tend to be more motivated to adopt smart technology, reflecting the role of progressive individuals in driving innovation and technology adoption. This finding aligns with the concept of intergenerational issues within family-owned businesses (Sharma et al., 1997), where newer generations often embody the “advanced management mindset” (Kellermanns et al., 2008). Conversely, a strong “conservatism” perspective within the business can pose a barrier to technology adoption, echoing the concerns raised by Miller et al. (2003) regarding resistance to change in family businesses.

Regarding the dynamics of family businesses, an intriguing paradox has been observed: Family businesses demonstrate a greater capacity (or discretion to act) to introduce innovations because of their inherent power and legitimacy, yet they exhibit a lower willingness to do so and are consequently less likely to adopt them (Chrisman et al., 2015). This capacity–willingness paradox appears to intensify as the family business transitions into later generations and power is distributed among more family members (Kraiczy et al., 2015). Our study provides empirical evidence in support of this notion. We identified an emergent theme of generational dynamics within family-owned businesses. One specific example from our findings is the concept of a “craftsman-like father block” (see Table A1), which falls under the subcategory of dynamics of family-owned businesses (Table 5). This concept presents a significant deterrent to the adoption of smart technologies. This finding offers a potential explanation for the paradox associated with transgenerational succession. Fathers who insist on traditional manual labor, rely on rules of thumb, and are unfamiliar with or resistant to new technologies can obstruct the introduction of smart technologies. This observation underscores the need for strategies addressing these generational dynamics to facilitate the adoption of technology.

On the other hand, the dynamics within family businesses are significantly influenced by the cultural and familial norms of their respective countries or regions. In Western countries, where farm consolidation and corporate management are more prevalent, these dynamics may not be as pronounced. However, in the Asia-Pacific region, particularly in countries such as China, Japan, Korea, Singapore, and Thailand, close family relationships, characterized by reciprocal family responsibilities known as "filial piety," are common (Phillips and Cheng, 2012). These familial expectations and cultural norms can impact technology adoption. Whereas, Phillips and Cheng (2012) point out that demographic changes and rapid socioeconomic development in the Asia-Pacific region have been linked to the decline of such familial relationships. As a result, even within the same country, there may be regional and generational differences in the dynamics of technology adoption. Therefore, examining the cultural and familial perspectives of a nation, as well as their evolution, in the context of smart technology adoption, presents an intriguing avenue for future research.

Our study's insights expand on family business theories by providing empirical evidence of the diverse perspectives within family-owned businesses and their influence on technology adoption decisions. The apparent influence of generational dynamics further enriches this theoretical framework, suggesting new areas of research. Given these unique dynamics and challenges, these findings underscore the need for tailored strategies promoting the adoption of technology in family-owned farms. Such strategies could include targeted educational initiatives, policy interventions that facilitate technology adoption, and support mechanisms to manage the intergenerational transition of business practices.

## 5.2. Practical implications

### 5.2.1. Insight for policymakers and agricultural support entities

The outcomes of our investigation underscore that individual farmer values and the overarching management policies of farms are pivotal components in the adoption of smart livestock technology. It is imperative for policymakers to acknowledge the sway of socio-technical elements, alongside industry currents such as public apprehensions regarding animal welfare, during the formulation of statutes to encourage the uptake of smart technology within the livestock farming sector.

The heterogeneity in agricultural policies across different livestock breeds may lead to disproportionate support and opportunities for adopting smart technology. Policymakers must recognize these disparities and strive to create more equitable landscape for farmers throughout the livestock sector.

Given the substantial influence of social relations and the impact of respected farmers within the community, there is a potential to harness these relationships in disseminating smart technology. Agricultural extension services could pinpoint and involve key influencers within farming communities to expedite technology adoption.

Structural impediments in the livestock industry, such as dwindling productivity and sector-wide contraction, can be tackled through tailored policy modifications and technological assistance. Policymakers and stakeholders should devise interventions that acknowledge the role of farm management policies and livestock industry trends in shaping the adoption of smart technology.

### 5.2.2. Recommendations for technology providers and agricultural advisory services

An appreciation of the sequential and iterative nature of the technology adoption process can influence the strategies of technology providers. Acknowledging that the drive to assimilate new technology can diminish without the opportunity for its introduction, vendors may need to explore adaptable and scalable solutions that can accommodate varying levels of investment and infrastructural capabilities.

Providers should also comprehend that the availability and compatibility of a technology with a farm's management system is a critical determinant of its assimilation. Consequently, offering comprehensive and customized implementation support could promote a positive appraisal and interpretation of technology by the farmers.

Our research underscores the significance of digital literacy, technological assistance, and technical experience in the adoption of smart livestock technologies. Vendors could enhance farmers' digital literacy, provide effective post-implementation assistance, and foster positive user experiences to boost adoption rates and ensure successful implementation.

Positive attitudes toward the adoption of smart technology can be nurtured through farmer training and peer-based learning. By disseminating success narratives and creating opportunities for farmers to learn from one another, vendors and advisory services can help establish a supportive community that encourages the adoption and sustained use of smart livestock technologies.

#### 5.2.3. Strategies for family-owned livestock farms

Understanding the dynamics of family-owned businesses and the tension between an "advanced management mindset" and "conservatism" is crucial in guiding family farms toward decisions related to technology adoption. It's worth noting that many family farms commonly prioritize workload reduction over improving management practices. Therefore, discussions regarding smart livestock technology adoption should emphasize how these technologies can alleviate workload and streamline farming practices, in addition to enhancing farm management.

Transparent dialogue about the perceived benefits and risks associated with smart technology, particularly in terms of potential workload reduction, can help manage intergenerational discord. It's essential to involve all family members in technology assessment and decision-making processes, demonstrating how these technologies can ease farming practices.

Resistance to change, especially from older family members, can impede technology adoption. Strategies such as targeted education and training for older family members, highlighting the benefits of workload reduction with new technologies, can help mitigate this resistance. Gradual introduction of new technologies can also alleviate perceived risks and complexities.

Our research indicates that factors influencing the adoption of smart livestock technology can vary, depending on the livestock species and farm management style. Therefore, family farms should customize their approach to meet the unique needs and contexts of different livestock species and farm management styles to maximize the effectiveness of technology adoption.

In succession planning, family farms need to consider the evolving nature of farming and the necessity of adopting smart technology for the farm's future survival and competitiveness. Emphasizing that smart livestock technology adoption can lead to both increased efficiency (and profitability) and reduced workloads can help present a win-win situation for all family members.

## 6. Conclusion

This study pioneers the utilization of an innovative methodology, fusing a scoping review, expert interviews, and an M-GTA, to delve into the complex dynamics of smart livestock technology adoption in Japan's unique agricultural context. It challenges and refines existing theories and models by providing a multilayered perspective on the factors that influence this process.

Our findings underscore the profound impact of individual farmer values, farm management policies, social relations, livestock industry trends, and agricultural policies on technology adoption. We also reveal the unique dynamics within family-owned businesses and the sequential and iterative nature of technology adoption. This process is notably influenced by technology availability, farmers' digital literacy, technology implementation support, and the observable effects of technology.

Additionally, by comparing the Japanese context with global contexts, we highlight the unique barriers and drivers in Japan, such as small-scale farming, strong government protection policies, and socio-cultural factors that influence technology adoption. This comparative analysis underscores the global relevance of our findings, suggesting that insights gained from Japan can inform strategies in other regions with both similar and different agricultural structures.

Despite the limitations of our study, which include a lack of direct engagement with farmers or policymakers, we believe our research offers robust findings with transferable insights. By providing an in-depth understanding of the context and comparing our results with existing literature, we enable readers to extrapolate our findings beyond Japan.

Looking ahead, future research should further investigate policy trends, the impact of social relations, and the dynamics of family-owned businesses across different cultural and socioeconomic contexts. Direct engagement with policymakers and farmers will be crucial in validating and expanding upon our findings.

In conclusion, our study not only lays a solid groundwork for future research on smart livestock technology adoption, but also bears the potential to inform policy decisions, guide technology providers, and influence management practices in family-owned livestock farms. Our findings particularly stress the need for managing the dynamic tension between "advanced management mindset" and "conservatism" within family-run farms. By offering these insights, we aim to stimulate further research and practice, ultimately promoting the successful adoption and utilization of smart livestock technologies in the agriculture industry.

## 7. CrediT authorship contribution statement

**Takumi Ohashi:** Conceptualization, Methodology, Resources, Formal analysis, Investigation, Writing – Original Draft, Visualization, Supervision, Project administration. **Miki Saijo:** Validation, Writing – Review & Editing. **Kento Suzuki:** Validation, Resources, Writing – Review & Editing, Funding acquisition. **Shinsuke Arafuka:** Validation, Writing – Review & Editing, Funding acquisition.

## 8. Acknowledgments

This research was conducted with the support of collaborative research funding from Eco-Pork Co. Ltd. The authors extend heartfelt gratitude to all participants who generously contributed their time and insights to this research. We also extend our sincere thanks to Mr. Dai Sakuma, whose invaluable counsel significantly shaped the qualitative methodology utilized in this study.

## 9. Declaration of AI and AI-assisted technologies in the writing process

During the preparation of this work, the authors used OpenAI's ChatGPT, an artificial intelligence language model, in order to streamline the initial drafting phase. After using the tool, the authors meticulously reviewed, edited, and supplemented the generated content as deemed necessary. Therefore, the authors assume full responsibility for the content of the publication.

## Appendix

Table A1 displays the worksheet developed during this study. It is important to note that all data collection and analysis were conducted in Japanese. The table provided here is the translated version in English, and it is worth considering that there may be a possibility of some loss in the original meaning during the translation process. Should readers require further clarification, the authors are available to provide additional information upon request.

**Table A1 Result of concept generation (translated).**

| No | Concept name | Definition | Theoretical memo | Variations |
|---|---|---|---|---|
| 1 | Adherence to social norms | A norm formed by sharing certain behaviors and values in a community, and that it is socially desirable to follow. | Some livestock farmers may find it desirable to consult the opinions of surrounding farmers and experts and not deviate significantly from existing methods. The opinions of surrounding farmers and the diffusion of technology may influence farmers' decisions to adopt technologies. | Influence from the surroundings. For example, it's hard to find one person in town who does something totally different. (I08_77) |
| 2 | Administrative impact | Impact of government policies and initiatives on the process of smart technology adoption by livestock farmers | Government action plays an important role in the process of smart technology adoption by livestock farmers. The government creates incentives for farmers to be willing to adopt by setting policies and rules to encourage technology diffusion and providing subsidies and grants. However, if farmers are skeptical about the effectiveness of the technology or are more concerned about the economic benefits, government efforts alone may not be enough to promote adoption. This suggests the importance of communication and collaboration between the government and farmers. | [name of association]. I consulted with the [name of association]. Generally, each prefectural livestock association is an organization that acts as a go-between for the subsidy menu. Well, they said, "How about this? So we installed a calving detection system called [product name]. (I01_6)<br>I think it's a thought that individuals may or may not want to put in. The point is, well, as a country, we are making such a move to try to introduce the device, but it is the people on the ground who will make the final decision. (External pressure, I10)<br>HACCP acquisition became important during the Olympics.<br>If the government sets the rules and the benefits are felt, farmers will do it. There is methane now, right? There are some feeds that can reduce methane. The government wants farmers to use such products. But the farmers don't want to use them. They say, "Well, how much money are we going to get from the methane reduction? That's why they don't want to do it and won't do it. So, if we say that we are going to pay for it, I think it will spread at once. Whether that is right or wrong is another story, but I think that if such subsidies are provided, farmers will actually take up the challenge. (Agri-environmental scheme membership, I10)<br>I have had discussions with people from [province name] and [province name] about grazing. Those two prefectures are pro-grazing prefectures, aren't they? In [prefecture name], they are doing a lot of support for grazing, so much so that people from outside the prefecture come to [prefecture name]. The prefecture subsidizes the grazing. (Management type, I10)<br>I think that [name of association], for example, is already trying to make their farms smarter, but in the end, it's all about the mechanism, isn't it? What I am doing, for example, is pasturage. [name of association] and I agree on the production of calves by pasturage, but I am working on the production of meat after pasturage, but there is a lot of pressure to sell that part.<br>It is difficult to make livestock farming smarter because of the structure of the cattle. It is not profitable to make marbled meat, and it is not profitable to milk cows. I don't think it will become popular because people are not willing to spend money on new technology when it is not profitable. If you introduce a new system, you have to include animal welfare, so even if you say, "Let the animals graze on pasture," the reality is that it is difficult for people to accept. If that happens, [name |
| | | | | of association] will say that it is no use doing so, and they will say that we should smarten the system to where we can get the money now.<br>For ordinary farmers, the agricultural cooperatives take care of various management tasks such as purchasing food and registering cattle, and they also provide subsidies. I think they have a very strong pipeline. (External pressure, I07) |
| 3 | Attention of the world | The phenomenon is more affected by the smart technology adoption process when there is contact with consumers, such as business management or farmers with a sixth industry. | With the shift to corporate management and the development of a sixth industry, livestock farmers and companies will need to strengthen their brands and communicate their image to consumers. This process may facilitate the introduction of smart technology. | Companies that sell their own brand of pork and other products are more likely to introduce<br>As the sixth industrialization progresses, we need to convey this image to consumers. Until now, B-to-C companies have done this, but now, if the 6th industrialization is to be carried out from production to distribution and sales all at once, it will be easier to introduce such a system, and I think it will be more effective (Agri-environmental scheme membership, I09). scheme membership, I09)<br>As I said before, egg-laying chickens are all-in, all-out. Like cows, there is no such thing as "animal welfare" in terms of how long the life span of a chicken can be prolonged, and in the end, chickens die within a few days. So, I think that the public's view of egg-laying hens is a very important factor, and since we are a corporation, we are very afraid that our corporate image will be threatened. That's why I think they are very afraid of their corporate image being threatened. When it comes to improving the image of the company, I think it will be easy to overcome the hurdle. (i05_41, 43, 45) |
| 4 | Availability of smart technology | Phenomena that indicate how easy or difficult it is to implement smart technologies, depending on livestock type, rearing environment, and produce | The feasibility of smart technology adoption in livestock farms is influenced by the product (milk, eggs, meat, etc.), the type of livestock (dairy cows, egg-laying hens, beef cattle, etc.), and the rearing environment (barns, grazing land, etc.). Dairy cows and egg-laying hens are easy to acquire biological data and are likely to benefit from smart technology. On the other hand, livestock raised on pastureland and livestock such as chickens, for which individual management is difficult, have difficulty adopting smart technologies. In addition, cost-effectiveness and business benefits are also considered as factors influencing the adoption process of smart technologies. | It seems to be relatively related. It has to do with how easy or hard it is to get biometric data.<br>I think there is a final product for each. Milk for dairy cows, eggs for egg-laying hens. Other products are meat. In other words, whether biological data can be obtained or not. For dairy cows and egg-laying hens, it is possible to know their health condition from their milk and eggs, but for meat, you cannot know unless you kill them. With meat, you don't know until you kill the animal, so I think there is a difference in how easy or difficult it is to obtain such biological data (Livestock breed, I02).<br>It depends on the cost per animal. For example, it is easy to introduce IoT for cattle and pigs, but it is impossible to manage individual chickens, so it is likely to be only environmental measurement. (Livestock breed, I03)<br>I think dairy cows could benefit from smart livestock production.<br>Egg-laying chickens, especially non-cage systems, are difficult (Livestock breed, I05)<br>Livestock kept in livestock barns are more acceptable. But depends on business benefit. Cost-effectiveness.<br>In a livestock barn, it is easy to see many things because the animals are kept in confinement. Temperature, electricity, and so on. First of all, the infrastructure is easy to use, and the scope of viewing and management is narrow. But it's not so easy because our grazing is so dynamic. (Livestock breed, I07) |
| 5 | Aversion to new things | Psychological hurdles to new smart technologies, especially among the elderly, making it difficult for them to adapt to and understand the technology | In the process of smart technology adoption by livestock farmers, older farmers tend to be resistant to new technologies and hesitant to adopt them. This resistance may stem from past experience, lack of knowledge, and anxiety about the unknown. | Like technology that is too new. For example, for old men, driving a car, for example, and automating it by themselves. The hurdle for them is that "I've never seen anything like this before," which is quite a psychological hurdle. I think there are a lot of people, especially the elderly, who don't want to adopt such a technology because they think it is too new and they don't understand what it is. (I03_26) |
| 6 | Awareness of the issues | The willingness and attitude to recognize and deal with one's own challenges and difficulties. | Problem-oriented livestock farmers understand their current situation and challenges and are willing to actively explore new technologies to seek solutions. Livestock farmers' self-perception (whether they are cattlemen or managers) and their vision for their work have a significant impact on the adoption | If you don't have the motivation to do something like this or to understand something through watching cows, it is difficult to motivate people to make the best use of what you tell them you can do. Therefore, I think that people who are in trouble or seeking solutions are more willing to get involved. So, as I said before, if a farmer does not look at cows properly, he will not be able to do anything with this technology, and he will fail. (I04_49)<br>I think it is important whether they have a vision or not, and whether they are interested in their work or not. I think that some people feel that their work is threatened by the fact that cattle technology is being used for smart agriculture. (I08_136) |

| | | | |
|---|---|---|---|
| | | of smart technologies. It is suggested that the willingness and success rate of technology adoption depends on the awareness of the problem and the qualities of the livestock farmer. | |
| 7 Breeding methods suitable for precision management | Impact on Smart Technology Adoption through Breeding Methods that Easily Achieve Precision Control | It has been suggested that husbandry practices suitable for precision management are important factors in the process of smart technology adoption by livestock farmers. Certain rearing systems (e.g., pig stalls and chicken cages) and artificial environments (e.g., windowless piggeries) tend to promote smart technology adoption because the benefits of technology adoption are clear. Farm size and management style are also relevant factors influencing smart technology adoption. Livestock type may also have an impact. | Dairy farming is more mechanized, automated, and easily introduced. Beef cattle specialize in calving monitoring. (Livestock breed, I01)<br>Currently 100 cows, in the future 1000 cows will have to rely on smart<br>Value per cow (Livestock breed, I04)<br>Large-scale management or combined farming, even in a full-time operation, will require smart technology. (Management type, I01)<br>I think it is easy to use a rearing system (stalls for pigs, cages for chickens) that can aim for super-efficiency in precision management. (Management type, I03)<br>In the case of dairy cows, the number of cows and the level of technology are increasing in the order of stanchions, tie stalls, and freestalls. Therefore, they are more active in this order (Management type, I04)<br>In pig production, herd size (number of animals per unit area) has an impact<br>I guess the larger the herd, the higher the production efficiency, but I think the management efficiency will take a toll. It depends on what you are managing, but I think you want to digitize that (Management type, I06)<br>Impact. It's easier to incorporate animal husbandry in the barns. (Management type, I07)<br>Easier to incorporate in an artificial environment because it is easier to see the effects (e.g., windowless pig barns) (Management type, I09)<br>Large size gives a positive image for introduction. Larger size gives a more positive image for introduction (Farm size, I02)<br>The larger the size, the easier it is to think about and introduce (Farm size, I03)<br>The larger the farm, the more eager (Farm size, I04)<br>I think milking will come in regardless of farm size. As a reduction of labor. I think so, especially in the area of milking cows. Even now, there are robot milkers for tethered cows, so I think the workload of the milking industry is huge. I think that technology will be introduced in this area, regardless of the scale of the operation. There is a shortage of human resources. (Farm size, I05)<br>The bigger the farm, the easier it will be to go smart. (Farm size, I07)<br>The larger the scale, the more effective it is (because you can visualize areas that are out of sight) (Farm size, I09)<br>For example, if you attach a sensor to each cow, it costs 10,000 to 30,000 yen, and the larger the scale, the more it costs. If the scale of the project is large, the amount of money will increase. Also, when it comes to detecting diseases and other problems, farmers, especially those who keep their animals tied up, are able to tell if something is wrong by looking at the animals every day. I don't think we should rely on a machine to do that. I think that the form of rearing will have an impact on the device that detects behavior. (I01_14) |
| 8 Challenges for new business and business expansion | Willingness to actively challenge new technologies and businesses and to break out of the status quo, and the practice of such | Innovative livestock farmers have the potential to improve the development of the livestock industry and its competitiveness in the market through the implementation of smart technologies. They have the potential to drive innovation in the livestock industry through sixth industrialization and new business models. Innovative livestock farmers may also be associated with factors such as | There are young farmers who are taking on challenges such as the sixth generation of production<br>Young farmers who have taken over the business are taking on challenges such as meat production (Entrepreneurship, I06)<br>Farmers who want to start a new business or break out of the status quo are more likely to adopt smart livestock production (Market knowledge, I09) |

| | | | | |
|---|---|---|---|---|
| | | | entrepreneurship and market knowledge in the livestock industry. It is also related to business succession to younger generations and age. Farmers who are open to new business and business expansion are more likely to adopt smart technology. | |
| 9 | Conflict between family business tradition and innovation | The tension that arises between livestock farmers' adherence to family management and inherited traditional methods and their attempts to innovate with new technologies and ideas | Farmers who stick to family-run or traditional methods tend to be reluctant to adopt new or smart technologies. On the other hand, farmers who have learned new knowledge and skills and have an entrepreneurial spirit may bring innovation to their farming practices. Education and background may influence this conflict and tension. Understanding how they are affected is important in the smart technology adoption process. | On the other hand, it may be difficult for a place that has been run by a family business.<br>I think I would stick to what has been passed down in the family. Maybe it also has something to do with educational background. Some of the people I have met say that they went to university to study economics and new things from the family business, so I can't make a blanket statement. For example, if an engineering person goes into agriculture, I feel that he or she has an entrepreneurial spirit. I think that people tend to inherit what their fathers have done and try to keep the tradition alive, but they tend not to take on new challenges. (Entrepreneurship, I03)<br>Traditionally, this system, some of the presidents I have seen have been doing things like this for a long time, and they maintain it, but when I became president, while doing the things I inherited from my father, I started to try new things, although most of them are inherited from my father, but I think I have to try new things a little bit. The timing of the change of generations is a little bit different in that I am trying things out a little bit. For example, if you listen to a conversation between a father and his son, they have completely different ways of thinking, and while the son is familiar with his own system, there are many points for improvement that he thinks should be done from a new perspective. I think the timing here is very important. (I03_55) |
| 10 | Constraints on land use | The adoption of smart technologies is affected by constraints such as land conditions, information infrastructure development, quarantine and environmental issues | A variety of factors influence the location of livestock farms, each of which influences the smart technology adoption process. The characteristics of the livestock industry (livestock breeds and feed supply networks) and local constraints (unknown land ownership and abandoned land issues) are among the main factors determining location conditions. The development of information infrastructure and limited radio frequency coverage are also important constraints in the adoption of smart technologies. These location constraints may be alleviated to a certain extent by the national and local governments through the development of information infrastructure and identification of land owners. | If it were pigs, there would be a lot of complaints about fecal matter, so the conditions have become such that it can only be done there.<br>We also need an information infrastructure, but we originally said it can only be done there. We can only do it there, so we want the information infrastructure there. You can't just move to a certain place. That's why I want the government to do its best (Location, I05)<br>(Cows and chickens) We can't do it unless we have a proper information infrastructure.<br>(Chicken) Considering the cost of transporting food, the distance from the port is important.<br>(Chicken and pig) There are restrictions on land conditions in terms of quarantine (Location, I05)<br>Many areas are outside the radio population coverage area due to odor and quarantine problems (Location, I06)<br>In [prefecture name], it seems that there is no more land. There is no land for abandoned farmland, even though it is said to be there. The owner of the land is unknown, so even at the prefectural level, there is no way to know who owns the land, so even if they wanted to lease it, they can't. The only land that remains, for example, is on a very steep slope by the seashore, and it is impossible to graze cattle there. (Management type, I10) |
| 11 | Corporate management perspective | From a corporate management perspective, production efficiency, profit, and environmental considerations should be considered important. | Since growth and the pursuit of profit are the goals of business management, they will move toward the introduction of smart technology if it is determined to be operationally positive. In contrast to family-owned sole proprietorships, new technologies are more likely to be used effectively in corporate management. Even for independent farmers who do not join a | There is an agricultural production corporation in Kagoshima with several dozen farmers making several hundred million yen. It's a very corporate organization. They give out all the information such as how many hours a person is involved in feeding a cow and how much he/she is paid per hour. And the selling price is about this much. There are also agricultural corporations that tell you how much profit they actually make after deducting this amount. Smart farmers, or those who have been in business for a long period of time, have a good sense of what is possible, so they are often interviewed. Then, they would say, "Well, let's give it a try. They say, "Well, if you use this, how much can you do? I think it's the labor saving and production efficiency. (Belief about the importance of human intervention, I07)<br>It's easier to introduce corporate pig farming than a sole proprietor of a family business. |
| | | | production cooperative, managerial decisions are particularly important. | After all, as a company, you have to grow, don't you? Sole proprietorships are fine as long as they can feed themselves. But I think that companies are more likely to introduce new technologies to improve their business performance. (Management type, I09)<br>Corporate pig farming is more effective because environmental considerations and the introduction of new technology are more likely to affect management (Agri-environmental scheme membership, I09).<br>From a business point of view, I think it is a business skill to be able to feel the issues in the first place. (I02_50)<br>For example, in terms of introduction, if the manager's perspective is enhanced, this will naturally have an impact, so in the end, I think that the farm manager's attitude toward management will have the greatest effect. I08_130)<br>I think that the use of smart livestock production is very effective in analyzing data and visualizing results, so I don't see the point of using it unless there is an awareness of the need to properly analyze the data and connect it to the next management step. (I09_18)<br>Do you have a sense of crisis?<br>With the cost of feed going up and more and more farmers going out of business, those who are seriously thinking about how to survive, those who have a sense of crisis, will take some kind of action, not only smart devices, but also new things. It's not so much that they want to make something big, but rather that they need to survive.<br>It's also a question of whether they are the kind of person who always has access to information.<br>We all have a sense of crisis, but if, for example, we decide to borrow money, is there someone around us we can talk to immediately? As I mentioned earlier, feed prices are skyrocketing, and there is a growing movement in each prefecture to seek subsidies from local governments. Individual farmers cannot take action on their own, and unless they can visualize the connection between the top management of the [name of association] and the prefectural governor, they will end up not taking action even after they have decided what to do. I guess. (Entrepreneurship, I10)<br>I think that the desire to work on something new will come when there is a sense of crisis, or when, for example, feed costs have risen to unprecedented levels and management is struggling. (I10_15) |
| 12 | Cost-effectiveness of technology adoption for different livestock species | Evaluate differences in adoption costs and adoption benefits for different livestock breeds when livestock farmers adopt smart technologies. | Different livestock breeds have different benefits and different considerations for recruitment costs and effectiveness.<br>Broilers are easy to trial and error, and it is easy to see effectiveness in short cycles.<br>Pigs and chickens are better managed and can be profitable with high turnover. Beef and dairy cattle, on the other hand, are less profitable and tend to be hesitant to adopt because of the high cost of the devices.<br>Dairy cattle and egg-laying chickens are considered easier to introduce because results are easy to obtain and the effects of introduction are easy to see. | However, if each type of livestock is not profitable, how about the cost performance? Pigs and chickens are managed and handled well, so turnover is good, and depending on the way it is done, there are places that make a profit, but again, it is easy to do because the product is small. In the case of meat, the money spent on both dairy and beef cattle is large, but the cost of feed and other expenses is quite small, and the profit is quite small. The profit margins are thin, and we operate on a bicycle. In reality, if you are going to charge 10,000 yen per cow for a device, it would be better to make a profit from that amount. (Livestock breed, I07)<br>In terms of livestock breeds, broilers are the easiest to trial and error. It takes 45 days to complete a cycle, which allows for five rotations per year. For pigs, it is about two or three cycles. Dairy cows are produced every day. (Livestock breed, I06)<br>Data is taken daily, dairy cows and egg-laying hens are easy to see the results, so it is easy to see the effects of introduction.<br>The ones for breeding are easy to introduce because they have a big impact on management (Livestock breed, I09) |
| 13 | Craftsman-like father Block | In family-run businesses, the phenomenon of fathers who insist on traditional manual labor and rules of thumb and are unfamiliar with new technologies hinder the introduction of smart technologies. | In family businesses, the father, as the decision maker, is highly influential in the smart technology adoption process. It is important that the father understands the technology and recognizes the need for it. However, if they have a commitment to traditional manual labor and rules of | Father. You, it's like you can't do without it. (I08_53, 55)<br>If the father didn't share that feeling, he probably wouldn't be introduced to it. (I09_28)<br>Exactly that of a family business. For example, if the son, the successor son, says that the father is a craftsman, then I think it would be difficult to introduce it. (I09_32)<br>In a company, it may be the president, or in a family business, it may be the father if he is the representative of the family. I think it is important for such decision makers to have such awareness. (I09_34) |
| | | | thumb, technology adoption may be difficult. | |
| 14 | Decline of the industry as a whole | Lack of vision among livestock farmers leads to business decline, a phenomenon that spreads throughout the industry | When livestock farmers have insufficient vision in the process of adopting smart technologies, management is only about matching numbers, and the result is decline. Clarity of vision is important because this decline affects the entire industry. A shared vision for the industry as a whole and the promotion of effective application of smart technology adoption could lead to sustainable development of the industry. | I have always felt that if there is no vision for management, it will just be a combination of numbers and decline. (I08_215) |
| 15 | Dependence on market prices | Shipment according to market price information, i.e., influenced by market prices rather than by one's own volition. | In some cases, it is difficult to ship according to market price information and to be actively aware of the market by oneself. For example, in the dairy industry, it is common to ship to [name of association] (Japan Agricultural Cooperative), and in some cases, the industry is far from being aware of the market as a business. Subordination to market prices may make it difficult for livestock producers to develop strategies to maximize competitiveness and profits. | (Cattle) Shipped to [name of association], far from being aware of the market as a business (Market knowledge, I05)<br>The farmer follows market price information. (Market knowledge, I07) |
| 16 | Difficulty in accessing technological information | Difficulty for farmers to obtain information on smart technologies | Farmers often experience difficulties in accessing technical information when adopting smart technologies. Issues with trade journals and the abundance of industry terminology are obstacles to information collection. Therefore, it is considered necessary for manufacturers and others to disseminate easy-to-understand information. In addition, in order to introduce smart technology, it is necessary to develop a mechanism for farmers to access new information.<br><br>On the other hand, the level of knowledge of farmers and their strong ties to entrepreneurship may not be a barrier for active farmers.<br>Relationships with universities and research institutions are also affected. | How little do farmers know about technology in the first place? (Access to technological information, I03)<br>I think that the difficulty in accessing technological information is limited to the swine industry. There are only four magazines in the swine industry. There are only four trade magazines. The problem with trade magazines is that they lack immediacy. There are only four monthly magazines. In addition, in the case of the pig industry, there are industry terms such as "breeding" and "semen," keywords that are not appropriate for the Internet. There are some companies that are open to the public and are doing YouTube videos, but it is difficult to make them available as contents on the Internet. Open and accessible to everyone (Access to technological information, I09)<br>Manufacturers do not provide easy-to-understand information. (e.g., how to use on Youtube, etc.) (Access to technological information, I02) |
| 17 | Difficulty in communicating new technologies | Lack of communication of new livestock-related technologies to farmers due to weak relationships between research institutions and farmers. | The transmission of new technical information on livestock production requires close involvement between the information provider, such as research institutions, and the recipient, the farmer. However, Japanese universities have few | Unlike overseas, Japanese universities, well, maybe only agricultural universities, do not have what is called "extension functions. If you go to overseas universities, there are farms attached to them. While there are farms attached to universities, there are also extension centers that are constantly disseminating information. After all, even if you do that, you will not be evaluated by Japanese universities. When we do experiments with farmers, we are told that it is a special case or that the sample size is not large enough. So, when the farmers are not able to say, "Please do that," it is |
| | | | extension functions, making it difficult for the information providers to disseminate information. In addition, in order for farmers to understand research information, it must be presented in an easy-to-understand manner. However, because it is difficult for farmers to easily access papers and researchers, there remain challenges in communicating information. Furthermore, researchers may not have a place where they can test new technologies, or there may be no one to connect farmers with research institutions. To solve these issues, it is important for the information provider to disseminate information in an easy-to-understand format and promote communication between farmers and researchers.<br><br>Strongly related to technical information that is difficult to access. | difficult to write a paper even if you do a demonstration test. (Access to technological information, I01)<br>How can we present research information in a concise manner? (Access to technological information, I04)<br>There are people from various universities and colleges, so how would you put it? It is quite troublesome for the presenter to create such a step-by-step process, but I think it is necessary to do so.<br>Farmers do not have easy access to papers and researchers (Access to technological information, I09)<br>There is no place for researchers to conduct tests (Access to technological information, I09)<br>No one to connect farmers with research institutions (Access to technological information, I09) |
| 18 | Difficulty in selecting competitive products | Phenomenon in which farmers are not clear on the difference in quality of competing products when selecting smart technology products, leading to confusion in their choice. | Farmers' inability to grasp the differences in quality and functionality, making it difficult for them to make a choice, is a barrier to the spread of smart agriculture. Clearly communicating the differentiation from competing products and their unique value could make it easier for customers to make a choice, and thus make smart agriculture more widespread. In addition, appropriate information should be provided, taking into account that strengths and weaknesses of each product exist. | The other day I heard someone who purchased our competitor's service and he said that the quality was totally different. He was using [product name], and he was using [product name], and the quality was totally different from [competitor's product name], and he said he felt [product name] was better. But if you look at it objectively from the outside, it is the same herd management system. But when you look at it objectively from the outside, it looks like the same herd management system or biological monitoring system, so people don't know the difference and buy it based on the specifications. He told me that he was very angry with me and told me that I should properly express the values that do not appear in the numerical specs. (I08_371, 373, 375, 377)<br>I think that if we can clarify this, smart agriculture will spread more widely, and I think there are strong and weak points in both. (I08_385)<br>That's right. It's hard to choose, so I don't know, so I think it's okay. (I08_395) |
| 19 | Discrepancy between animal welfare compliance and smart technology | Phenomenon in livestock farmers that as animal welfare compliance increases, it becomes more difficult to adopt and apply smart technologies. | The relationship between animal welfare compliance and the adoption of smart technologies differs depending on the livestock species. In the case of chickens, non-cage keeping is suitable for animal welfare, but it causes a gap with smart agriculture. For cattle and pigs, IoT adoption is difficult when they must be herded instead of individually, but IoT is easy to adopt under certain circumstances (e.g., single stall, cage-keeping). In the process of smart technology adoption by livestock farmers, it is necessary to understand the divergence between animal welfare and smart technology and consider appropriate technology adoption for each livestock species. | I think egg-laying chickens will go against animal welfare in the future. Now there are non-cage, cage-free systems, such as Aviary and flat-feeding. I think the first two are cage-keeping and non-cage-keeping. After all, it is easier to manage chickens in cages, isn't it? They don't move. However, in the case of "Aviary" and "Flat-feeding," the chickens are free to do as they please, but they tend to lay eggs here and there, and it is difficult to find dead chickens. I think that the more we try to be animal welfare-oriented in egg-laying chickens, the more we try to be cage-free, the more there is a gap with smart agriculture. (Livestock breed, I05)<br>I guess it depends a little bit on the livestock breed. Cattle are also involved in animal welfare, but when it comes to welfare, I have the impression that it will be difficult to introduce IoT if we have to go from individual chickens or pigs to a herd. On the other hand, however, IoT is very easy to implement in the case of stalling pigs or keeping chickens in cages, for example, by patrolling the area or detecting the estrus of cows in the same place. It's like putting a machine in a place where it was being done by a person. If you let them go, it's hard to understand. (Management type, I03)<br>In the case of egg-laying hens, do you have expectations for both large scale farming and animal welfare? (I05_14) |
| 20 | Durability of technology | Ability of equipment to maintain functionality under a variety of conditions and withstand adverse environments and shocks | Equipment durability and environmental adaptability are key factors in the process of smart technology adoption by livestock farmers. The challenges inherent in the livestock environment, such as high ammonia concentrations, drops, and wind and rain effects, can easily lead to equipment failure, and the equipment must be able to handle these conditions. | The equipment must be difficult to wear out and simple to operate.<br>Should I write "durability" because it is easy to break down if ammonia concentration is high or if it is dropped depending on where people use it? If it is installed at a fixed point in a cultivated area, it will be exposed to wind and rain, and it will be ruined. Since there is always the possibility of dropping the terminal as it moves, it must be protected from the shock of being dropped, and it must also be able to operate in ammonia, dust, and poor environments. Basically, the reason PCs were not introduced in livestock production was because they were spindle type with a fan turning. They were equipped with a refrigerant system, which would ruin them quickly. (Perceived ease of use, I06) |
| 21 | Ease of installation | Easy to install and operate and less labor intensive for livestock farmers to adopt smart technologies | Ease of technology installation and operation is important for livestock farmers when adopting smart technologies. Value-added features such as veterinary installation of the technology and health checks of cows could also be a deciding factor in adoption. If technology that can be easily switched to current work methods is developed, adoption may be promoted even among those who are not comfortable with new technology. | I think that the main motivation is that it would be troublesome to add something new. (I01_8)<br>It's not a hassle, right? Labor saving. Or rather, I want to do it easily. For example, the [product name xxx] I mentioned earlier is installed by a veterinarian, who attaches it to the cow. So I don't want to do it myself. And if the veterinarian comes, he can check the condition of the cows along with the installation of the [product name xxxx]. They could attach it to something in the process. I think that kind of thing is a big part of it. (I01_14)<br>It is very important to be able to do this easily, and it takes a lot of labor for farmers to put something on cows that is now called "wearable. Even with the [name of product] that I went to see, there is a rule that it must be worn snugly around the neck, and the sensor must be behind the ear. (I04_26)<br>I'm not sure if there is such a thing that can be easily switched from the current product to make it easier for people to use, but if there is such a thing that can be easily switched without any difference from the current product, it would be important not only for people who like new things, but also for a wider range of people to use it. I think it is important to have something that can be easily switched without any difference from the current one, so that not only people who like new things but also various people can use it. (I10_25) |
| 22 | Ease of use of technology | That ease of use, visibility, and less hassle are important factors when adopting smart technology, in addition to device performance. | The smart technology adoption process among livestock farmers suggests that ease of use and visibility are deciding factors in adoption. Especially for the elderly, technology that is easy to use and less cumbersome is more acceptable. The background and characteristics of individual livestock farmers may also influence the technology adoption decision, as "sociodemographic attributes" are relevant. | Ease of use and visibility as well as performance of smart devices matter. (Perceived ease of use, I01)<br>Greatly relevant. If you don't think it's easy to use, you won't adopt it.<br>It has to do with "sociodemographic attributes" or something like that. (Perceived ease of use, I02)<br>I think it is important. Especially for the elderly. (Perceived ease of use, I03)<br>Of course I think so (Perceived ease of use, I04)<br>Perceived ease of use, I03 (Perceived ease of use, I07)<br>Less hassle is important (Perceived ease of use, I10)<br>If there is some good technology, for example, a smartphone like the RakuRakuPhone, and if we can go to a place like an exhibition for livestock breeding, for example, and see that it works well at all when we touch it (I03_28) |
| 23 | Expectations for ability expansion | Expectations for the use of smart technology to complement and extend human capabilities in early detection of signs of heat, calving, and disease that livestock farmers often miss with traditional methods | Smart technology could extend human capabilities and enable livestock farmers to manage more efficiently and with greater precision, and improve knowledge transfer and accuracy of situational judgment among employees. | In the case of cows, the earlier we can detect signs of estrus, calving, and disease that cannot be seen with the human eye, the more money we can make, so I think that is the second point (I04_6).<br>The second point is to cultivate the eyes. Even if you tell a new employee why you think it's time for shipment, they cannot understand it, but you can easily check the answer by checking it with a computer, I think this will contribute to the expansion of human capabilities. I wonder if there will be qualitative or quantitative effects. (I06_23) |
| 24 | Expectations for formalizing tacit knowledge | Expectations of using smart technology to transform tacit knowledge into formal knowledge to improve efficiency and productivity | Some farmers expect smart technology to replace decision criteria that have traditionally relied on intuition and experience with objective data. For example, there are indications that smart technology is contributing to increased efficiency and productivity in shipping decisions, weight measurement, and temperature and humidity control. | There is demand for technology to formalize tacit knowledge.<br>Like [product name], I'd like to measure [product name] to see why I made the shipping decision now, and see if it's because it was this number. With a scale, you can't check it. On a very large farm, you could spend an entire day just moving the scale around.<br>Temperature and humidity are also important. When we were raising broilers and such, we noticed that it was getting colder and colder. If we didn't close the windows, we would have to open the windows to ventilate the broilers because the $CO_2$ concentration would increase and the broilers' growth would be delayed. But by installing a temperature sensor, he said, it was good to be able to |
| | | | | know what time he made that action. It was a good thing that we added a digital sensor (Perceived value, I06). |
| 25 | Expectations for labor saving | Expectations for the benefits that smart technology can bring to livestock farmers in terms of reduced labor and increased efficiency | Labor savings is one of the major attractions for livestock farmers in adopting smart technologies, and they tend to positively accept such technologies when the benefits are clear. Technologies such as automatic feeders, which allow farmers to realize labor savings, are well received by livestock farmers. Furthermore, with the aging of the population, labor savings will become an important factor as more and more situations require labor reduction. Therefore, clarifying labor-saving effects in smart technology development and dissemination may increase the willingness of livestock farmers to adopt such technologies. | Labor saving is nice. (Perceived usefulness, I07)<br>I'm currently using a system like an automatic feeder that feeds my dogs when I press a button from my smart phone from home. That kind of thing is very easy. They are very pleased with the labor savings. (Perceived usefulness, I07)<br>They are willing to accept it as long as it contributes to labor savings. (Belief about the importance of human intervention, I07)<br>If anything, labor saving is more important when the population ages. The aging of the population is a big factor in labor saving, because people who eat from cows can still eat to a certain extent. If anything, they are looking for smart, labor-saving products. (Entrepreneurship, I07) |
| 26 | Expectations for margin creation | Expectations of creating time and psychological leeway by reducing hazardous and demanding work | Labor-saving benefits are a major motivation for livestock farmers when adopting smart technologies. The more clearly these benefits are perceived, the more farmers appreciate the value of the technology and are more willing to adopt it. In addition, the time margin created by labor reduction and productivity improvement is an important perspective for farmers, and is a factor that raises their expectations for technologies that will show these benefits. In addition, increased safety and profits from reduced hazardous and arduous work are also important factors in the smart technology adoption process.<br>Time and mental capacity will lead to new challenges. | I thought it was important from the standpoint of creating a time margin. (Perceived value, I03)<br>(Perceived value, I04)<br>(Cattle) Labor saving effect is high (Perceived value, I05)<br>The value is felt when productivity gains and labor savings are perceived. Again, it would be production efficiency and labor saving. Since it is hard, dirty, and dangerous, well, that kind of dangerous work and hard labor will be reduced. I think that will be a major factor in reducing the amount of labor. That's what I'm thinking of.<br>In the end, I heard that profits increased even a little after this was done. Or, "It's easier now that I've added this. That is the most valuable thing, isn't it? (Perceived value, I07)<br>I think the biggest thing is the reduction of labor. (I04_2)<br>From a dairy farmer's standpoint, it's really a labor reduction. I really think that's the biggest thing, and from the farmer's standpoint, perspective, and to improve the farmer's farmer welfare, it's labor load reduction. (I05_14)<br>I think that when farmers can have time to do things with their families, and when they can have three times more time than before, they will be very active. (I05_21)<br>I developed [product name] simply because I wanted to save labor by throwing pigs one by one onto a scale, and since each pig weighs about 120 kg, I wanted to find them out of a large herd and put them on the scale, I think one of the motivations of farmers is to solve that, to solve that. (I06_2)<br>If we can save labor, we will have more time, which will give us more room in our minds to try out new approaches, or in other words, we will have more time to do various things. (I06_14) |
| 27 | Expectations for problem solving | Expectations about the benefits of introducing smart technology to solve problems, improve efficiency, and share information | The deciding factor for livestock farmers to adopt smart technology is largely due to the expectation that it will solve the problems they perceive in their own companies. Some farmers face a variety of challenges, such as reducing the time and effort required for record keeping, refining data analysis, and making it easier to share information. | The deciding factor is when the farmers think that they can solve their own problems, problems that they feel in their own companies, well, of course (I02_26)<br>Yes, that's right. That's a killer word. Let's see. Well, it may be a detailed explanation of the product, but, for example, our product can reduce the time and effort required to keep records, such as ledgers. (I02_28)<br>For example, our product enables more precise data analysis and visualization of farm conditions, and also enables the sharing of information among employees. And, depending on the farmer, many farmers feel that this is an issue, so if we tell them that our system can do this and solve it, and if we show them the data, they may decide to introduce the system. (I02_30) |
| 28 | Expectations for productivity improvement | Expectations of productivity gains in terms of economic efficiency, labor savings, labor savings, and increased profits | For livestock farmers, the value of adopting smart technology is clearly cost-effective, and factors such as economic efficiency, production efficiency, labor savings, reduction of hazardous work, and increased profits drive adoption. Different farmers may value these factors to different degrees, and their perception of value when considering introduction is individualized.<br>In addition, post-introduction value perceptions may influence livestock farmers' continued adoption of the technology and dissemination of the technology to other farmers. | Stand-alone implementation will not be implemented unless cost-effectiveness is clear. (Perceived value, I01)<br>(Chicken) Value as economic efficiency (Perceived value, I05)<br>Value is perceived when productivity and labor savings are perceived.<br>Again, it would be production efficiency and labor saving. Since it is hard, dirty, and dangerous, well, that kind of dangerous work and hard labor will be reduced. I think that will be a major factor in reducing the amount of labor. That's what I'm thinking of.<br>In the end, I heard that profits increased even a little after this was done. Or, "It's easier now that I've added this. That is the most valuable thing, isn't it? (Perceived value, I07)<br>After all, it's money.<br>For example, many people who have time to use devices think about how they can increase their sales. I wonder if there are people who put in devices for the purpose of creating leisure time. (Perceived value, I10)<br>I think that the essence of any industry is to save labor and increase productivity per hour, and I think that is the point in agriculture. (I06_4)<br>I think the most important motivation for the introduction of the system is to find out the percentage of increase in sales from the current level of productivity. (I10_8) |
| 29 | Facility renewal opportunities | Facilitates the introduction of smart technology when facilities are rebuilt or expanded due to aging or increased head count, etc. | When new or renovated barns are built or renovated, farmers are more likely to consider the introduction of new smart technologies. Taking advantage of this opportunity, the introduction of new technology could improve farmers' management efficiency and productivity. This concept is important to understand the timing and factors that may influence a farmer's adoption of a new technology. Consideration of market trends and investment affordability are essential for updating facilities through increased head counts. | A budget line is essential. For example, building a new piggery is a good start.<br>Building a new piggery is basically a matter of durability, for example, 30 years, so rebuilding it once would be a good start. That kind of building durability is one thing, but there are also other things, such as increasing the number of pigs, for example (Perceived room for investment, I09)<br>If the livestock barn is renovated or newly built, wouldn't they be introduced incidentally? (e.g., cow brushes) (Perceived value, I01)<br>Not only physical renovation or new construction of barns, but also expansion of the scale of barns would be one of the opportunities. (I01_6)<br>In the case of corporate pig farming, for example, building a new farm or rebuilding a piggery, as I mentioned earlier, I think such management timing is significant. (I09_39) |
| 30 | Familiarity with smart technology | Younger generations and those with higher levels of education have a more accepting attitude toward smart technology | Younger generations are more likely to accept smart technology because they are more digitally literate and comfortable with smartphones and apps. Those with higher levels of education are also more flexible with new technology and often find smart technology useful. These people tend to be more familiar with smart technology and more willing to adopt it. Age and education level are important factors that influence affinity for smart technology. They are also influenced by their daily information gathering. | Younger generations are more likely to accept smart technology. (Socio-demographic characteristics, I01)<br>Younger people are more, what do you call it, digital-divide against digital. The younger you are, the more digitally literate you are (Socio-demographic characteristics, I06)<br>I would assume that younger people are more likely to use, you know, smartphone apps, and they are more likely to accept it, and it's true that people with higher levels of education are more likely to accept it. (Socio-demographic characteristics, I07)<br>I think it would affect people who gather all kinds of knowledge on a regular basis, simply that smart is useful. I would like to use it. (IT knowledge, I07)<br>I think there is also this. In particular, I think it depends on how much you have been involved with smartphones and such things from a young age. (IT knowledge, I03)<br>Yes. It is close in meaning to age and market knowledge (IT knowledge, I02)<br>I think it's the degree of familiarity with smart phones, PCs, and PCs. There are a lot of farmers who are not interested in smart phones and PCs. (I07_42, 44) |
| 31 | Farmers' beliefs about breeding methods | Farmers' beliefs and commitment that influence smart technology adoption | Some farmers have a variety of beliefs and preoccupations, such as those who believe their way is best, while others are more concerned with animal welfare and environmental conservation. Farmers with | The older the generation, the more likely they are to say that their own way is the best way and that it does not matter if such a guideline is established (Socio-demographic characteristics, I01).<br>The older the generation, the more likely they are to say that their own way is the best and that it doesn't matter if such guidelines are established. (Belief about the importance of human intervention, I01) |
| | | | strong beliefs often consider adopting smart technology based on their own beliefs and use technology as an adjunct. On the other hand, farmers who are more conscious of animal welfare and environmental conservation tend to be more accepting of smart technology. The strength and direction of beliefs are factors that influence acceptance of smart technologies.<br>Older and smaller farmers tend to have stronger beliefs. There is a perception that technology is supportive of human intervention. | People who treat animals as living creatures rather than industrial animals are more likely to have such beliefs.<br>Some people are like that. (Belief about the importance of human intervention, I02)<br>I feel that farmers, especially small farmers, have stronger commitment and beliefs. (Belief about importance of human intervention, I03)<br>People who can see animals without smart technology can use smart technology effectively.<br>It's people like this who succeed. I have seen people who have beliefs and say, "I think so, but if a machine can do it, let's try it," and I agree with them.<br>You can't be too particular about your own beliefs, but if you understand that you can do human intervention through smart technology, that's strong (Belief about the importance of human intervention, I04)<br>The belief part is where you look at the cows and make a decision, and that's where you don't let them intervene. In the end, we don't let them intervene in the final decision, but we do let them make the final decision, but first of all, it's a lead in the system, or make the lead more efficient, and so on. When you have to do this job, do this job, and do this job, you can make it easier by using information. (Belief about the importance of human intervention, I07)<br>Many people believe that human intervention is essential and that technology is an adjunct to it (Belief about the importance of human intervention, I09)<br>Do you care about AW, or circulation and so on = more likely to include people who are more conscious?<br>I think that people who are concerned about animal welfare, environmental conservation, circulation, etc. are more likely to be smart enough to join. People who are less conscious of such things are less likely to join. (Others, I07)<br>The direct effect on milk production. Calving is also directly related to production, whether the calves are stillborn or not. Even illnesses, to put it bluntly, have a direct impact on production, but even if a cow has a slight fever, it is not as if the milk yield will decrease immediately. It is possible that the difference in motivation is influenced by the difference in sensation. (I01_16)<br>I don't use devices because I can't train people.<br>One example I know is that people have to watch them. We have 10,000 head of cattle, and one person has to watch 400 head. In the end, when you put in a device, you can't train people. They lose the ability to look after the cows. That's why they don't put in devices. Because you are responsible for watching the cows every day with your own eyes, you can learn, you can notice, and you can notice even subtle changes. I think one of the purposes of the device is to eliminate the differences so that everyone can notice such subtle changes. (Belief about the importance of human intervention, I10) |
| 32 | Fit with on-site routine | The impact of smart technology adoption on whether smart technology fits into the current workflow. | Perceived suitability is an important factor in livestock farmers' smart technology adoption process. Technologies that are perceived to be highly compatible with current work flows are likely to increase adoption intentions. | Yes, very much so. Need to consider whether it fits into the workflow at the farm (Perceived value, I02)<br>Interfaces and methods of use that understand field work are required (Perceived ease of use, I09)<br>Easier to implement if no additional work is required (Perceived value, I09) |
| 33 | Impact of financial support | Impact and significance of grants, subsidies, loans, and other financial support in adopting smart technologies | Financial support plays an important role in smart technology adoption, and many farmers tend to experiment with technologies because they feel they are good value for money. The presence of subsidies or grants is often a deciding factor in technology adoption, especially when the initial investment is large, and many farmers find it difficult to take the | There is a great deal of relationship. Essential to spreading the word. Many people think, "I'll give it a try because it's a good deal."<br>There are a lot of people who say, "I'm going to borrow it, or rather, I'm going to subsidize it for farmers. So, I'll try it because it's there. (Access to financial support/loan, I02)<br>For example, is it good to have something where you can try it with a small amount at first, and if it works out, you can borrow a lot?<br>(Access to financial support/loan, I02). Of course, if it pays off, I would borrow, but when I don't know how it works in practice, it would be good to have a system where I can try a small amount at first, and if I like it, I can go on to do a lot more (Access to financial support/loan, I03) financial |
| | | | plunge without significant subsidies. In addition, the amount of subsidies varies by livestock type, and these differences also affect differences in technology adoption. Accessibility to information and linkage to investment affordability also need to be considered. | support/loan, I03)<br>I think it is difficult to take the plunge without a big subsidy, such as a half-price subsidy (Access to financial support/loan, I04)<br>Chickens, too, are basically independent and profitable, but if subsidies are available, I would do it already. (Access to financial support/loan, I04)<br>Hog farming is very much borrowing or self-financing.<br>Cattle are very heavily subsidized because there are so many farmers, so from MAFF's point of view, it's a vote winner. So we need to appeal to the people involved in pig farming as well, and if the number of farmers is less than 4,000, we need to increase the number of people working on the farms and firmly appeal to the LDP that it is one vote-getter. (Access to financial support/loan, I06)<br>Need to provide information on subsidies and grants (Access to financial support/loan, I09)<br>Linked to investment affordability<br>Smart devices, costly. Putting 10 cows and 20 cows in a larger scale is honestly no different (Access to financial support/loan, I10)<br>I think 70% of the smart farming was cattle, and it was the same with Corona, but pigs and chickens go around in circles, but they were originally started on a small scale where they could not get subsidies, and then they borrowed money from banks to expand, and they got financial support from feed manufacturers and so on. I think some people think that we don't need subsidies because we started out on a small scale with no subsidies, and then we borrowed money from banks to expand and received financial support from feed makers. (I06_33)<br>Now that subsidies are actually being given because of the high cost of feed, I think they want to make some changes at that time. (I10_19) |
| 34 | Importance of face-to-face communication | The concept that face-to-face communication with livestock farmers is a prerequisite for building trust and business. | Face-to-face communication is important in livestock farmers' attitudes, which may influence their evaluation of service quality. While some livestock farmers believe that face to face contact is unimportant, the majority view face-to-face communication as important. This assumption of communication may be a hindrance to progress in the industry as a whole. Changing the way manufacturers and producers interact may be necessary to advance the industry. | This is about the mindset of farmers. They have a premise that companies that don't come in person are no good. So, recently, they're saying that [product name]'s service has really deteriorated. When we dig deeper, it's like, "Why don't they come to greet us each season?" The service itself wasn't a problem at all. So, what's the issue? They say things like, "If you're passing by, at least leave your business card." I thought only a few people would say that, but actually, the people we were involved with didn't care about that at all. Most of them did care about such things; the majority wanted them to show their faces. I think if we don't change the way manufacturers and producers interact, this industry won't be able to accelerate in the first place. (I08_414, 416, 418, 420, 422)<br>If we allocate people to such tasks and the company goes bankrupt, it's meaningless in the first place. (I08_426) |
| 35 | Inadequate information infrastructure | Challenges and constraints for livestock farmers to adopt smart technologies when they have inadequate information infrastructure such as internet connectivity and radio frequency environment | There are farms and regions with inadequate Internet access, which may make the use of smart technology difficult. On the other hand, odor, environmental, and land issues may force some farmers to set up farms in such areas. To address this, technologies that can be used offline and local data processing are required. | There are many services that cannot be used without an Internet connection. There are many services that cannot be used without internet access. Also, some farms, such as [company name], do not have internet access. (Location, I02)<br>Network problem. People who don't have internet access can't use IoT.<br>It would be nice if it was rechargeable in a stand-alone kind of way. It is difficult to connect to the internet and fly Wifi to see it on a smartphone. On the other hand, I have to go deep into the mountains in Japan to see it. (Location, I03)<br>Smart Farms are still the way to go. I think it is absolutely necessary to have a proper infrastructure and information infrastructure. Cows, pigs, chickens, etc. (Location, I05)<br>That's why, in the development stage of [product name], there are many areas outside of the radio population coverage area. So we heard from [company name] that in order to get an optical fiber line, there must be 10 households or convenience stores in the area. And since there is no fixed line, we had to install a local processing system. The update of [product name] was done in the town at the foot of the mountain. If used, the system had to be able to be used offline. (Location, I06)<br>It is difficult to use IoT, etc. because the project is in a mountainous area, a marginal village, etc. (Location, I07) |
| | | | | Internet environment is essential, so it is not suitable for places where it is not easy to develop it (Location, I09) |
| 36 | Income not proportional to effort | Situation where livestock farmers feel that their efforts have no direct impact on their bottom line. | The lack of correlation between effort and revenue can be a significant barrier in the smart technology adoption process for livestock farmers. Farmers may be reluctant to adopt new technologies and initiatives because they feel that their efforts have no impact on their earnings. This may lead to resistance to smart technology adoption and inability to adapt to change. Lack of conviction may also contribute to farmers' unwillingness to adopt technology, creating a situation where they feel there is no room for effort. | If we talk about the basic issues, as I mentioned earlier, there is the assumption that the agricultural cooperatives will purchase the entire amount of milk. And no matter how much effort you make, the price of milk is not decided by you, but by the government or by a designated organization. It is important to reduce the number of sick cows, but that is all there is to it. (I08_99)<br>I think that it is no good to do something like that, so I don't think that new things will come in, so I feel that, in [the interviewer's] expression, there is not enough space. (I08_104) |
| 37 | Inertia | A phenomenon in which farmers accustomed to the current feeding system and management style find it difficult to transition to new technologies and smart farming. | Farmers' attitudes and management styles may differ in their acceptance of smart farming technology. Farmers who are satisfied with the status quo or who have their hands full maintaining already established systems may find the transition to new technologies difficult. This inertia can be a factor affecting the diffusion and adoption process of smart agriculture. | Conscious farmers are more likely to accept. Those who keep the animals in the current system are less likely to enter. (Agri-environmental scheme membership, I07)<br>It's difficult to change the gears because the gears have already turned. I understand that. It's as if we can't stop cycling now. Unless we have a lot of time to spare. (Agri-environmental scheme membership, I07) |
| 38 | Influence from influential producers | Impact of referrals and evaluations from trusted people in the community and production network | If an influential producer in a region or production network tries a new technology and finds it effective, the information is considered reliable and is more likely to be disseminated to other producers. Conversely, if a new technology is tried and found to be ineffective, that information will also spread, making dissemination difficult. It is suggested that joint research with influential producers and successful examples from practice are important to promote dissemination.<br>With small-scale farmers, regional ties are particularly strong. | It is important to show successful examples by practicing, and not just selling what you've learned from others.<br>I don't think there are that many examples because I think people who are playing a central role are trying things out for themselves and introducing them to others, not selling them to others (Production degree centrality, I04).<br>I think there is this too. Especially with small farmers, there seems to be a large number of networks, so this may be important in such areas. (Production degree centrality, I03)<br>For example, in the case of arable cultivation, the area determines who is responsible for watering the land, and in the case of rice cultivation, who is responsible for watering the land. For example, in the case of rice cultivation, it is decided who is in charge of watering the fields, and this person is the first one to water the fields. It's like if it's equipment that was put in there, it must be good. In the case of pig farming, there are pig farming groups, aren't there? They purchase feed jointly. In other words, if the information is considered valid within the group, it will spread to several companies and farmers at once. In terms of promoting the spread and expansion, I feel that groups cannot be ignored. (Production degree centrality, I06)<br>There is a central person everywhere. The boss. If that person says we can use it, we will use it too. The boss is important. It is better to collaborate with the boss. The reverse is also true: if someone says, "We can't use this," it will spread at once. (Production degree centrality, I07)<br>Influence from a predecessor or influential producer (Production degree centrality, I08) |
| 39 | Influence of word-of-mouth on decision making | Influence of external factors such as word of mouth by veterinarians, feed companies, and other farmers on decision making in the smart technology adoption process | Word of mouth among livestock farmers can be very influential in smart technology adoption. In addition, the opinions of veterinarians can become the industry standard, and information provided by breeding companies can influence management. | I think the trend is well, especially in the main business. Did you say last week that it is now less than 4,000 farmers and 3,700? In such a small area, there are many people with loud voices. I think that the number of farmers who are vocal in this small industry is increasing, and what veterinarians and others say will become the standard in the industry. (Access to technological information, I06)<br>I think the same applies to all livestock, but the amount spent on feed is so large that it affects the management of the livestock business, and the feed cost is particularly high for cattle, so there is a lot of contact with feed companies in order to get information from them, to improve efficiency, and to |
| | | | Word of mouth can work both negatively and positively. | consider such management. (Access to technological information, I05)<br>Word of mouth is the most influential (Production degree centrality, I06)<br>There is word-of-mouth pressure (fear of being out of fashion) (External pressure, I02)<br>Word of mouth among farmers is highly effective (External pressure, I09)<br>Neighbors, right? Sometimes it's like, "Oh, their results are good, and we use the same kind of pigs in our neighborhood, but our results are bad, why? (I02_63) I think that the local community is the most important factor in dairy farming. The most trustworthy thing is the community. I think the most important thing is that it is spread by someone else's introduction. (I05_29)<br>For example, if it is a veterinarian, a food delivery person, or a manufacturer, someone who can connect with the outside world would be a great help. (I09_41) |
| 40 | Information disconnection | Situation where livestock farmers are unaware of the existence and benefits of smart technology and have limited information exchange with the outside world | Livestock farmers do not have much contact with the outside world and may be disconnected from new information, which may slow the adoption of smart technologies. | It's like there are so many useful things in the world, but you don't know it and you're disconnected. (I08_89)<br>Maybe it's like we are disconnected from the outside world. (I08_97) |
| 41 | Information-gathering ability | Ability to actively gather limited information | Those who are better able to gather information tend to be more proactive in adopting new technologies. Younger generations in particular are more familiar with digital technology, making it easier for them to gather information. The ability to gather information on costs and knowledge of the market are also important, and the more knowledgeable a person is, the more positive he or she may be about adopting technology. Willingness to gather information and communication skills are also important factors, and people with higher levels of these skills are more likely to introduce new technologies. Younger workers are more likely to possess advanced knowledge. | In other words, they must be able to go out and get information on their own. In other words, they are those who are willing to go out and get information on their own, but they need to communicate in order to do so, and they are willing to do so without being heavy on their feet. (Access to technological information, I10)<br>There are a lot of presidents in their 30s, who are in the second generation after the first generation has ended. Well, in that sense, they are able to use IT and gather information. (Access to technological information, I06)<br>For example, young producers are using digital technology to gather information. (Access to technological information, I09)<br>Ability to gather information on costs, market knowledge is necessary (Market knowledge, I04)<br>Knowledgeable people are willing to introduce the system (Market knowledge, I02)<br>Willingness to gather information. (Socio-demographic characteristics, I04)<br>The next generation has more advanced knowledge (Market knowledge, I06)<br>I guess it is a matter of motivation, but if there are places where [name of association], agricultural newspapers, or local administrators make and distribute flyers, they can get information from those places, or if the farmers are very advanced, they can always search the Internet. If the farmers are very advanced, there are always people looking for information on the Internet. (I07_28) |
| 42 | Information gathering from industry publications | How to obtain smart technology and livestock-related information through specialized trade journals | Trade and professional journals are important sources of information in the livestock industry. In gathering information, trade and professional journals provide access to new technologies and knowledge. However, the frequency and variety of updates are limited depending on the type of livestock. | Knowledge of smart technology may be obtained from trade journals, etc. Is it abundant in dairy farmers? If they are introduced in trade magazines, etc., they can be exposed to such technological information. (Market knowledge, I01)<br>I think there are about 2,000 trade magazines in Japan. However, there are still some fixed markets in the agriculture, forestry, fisheries, construction, and food industries, and within the agriculture, forestry, and fisheries industry, there is the livestock and fisheries industry. Cattle and chickens have daily or bi-weekly magazines, but pigs only have monthly magazines, so there are only four magazines in total. (Access to technological information, I06)<br>Information on pig farming through information booklets (External pressure, I09) |
| 43 | Interest in living things | Farmers should take an interest in their livestock as living creatures, not as industrial animals. | Farmers' interests and beliefs about their animals influence their ability to maximize the effectiveness of smart technologies. Farmers who care about their animals and value each animal's life are more likely to make effective use of smart technology. These farmers are interested in using technology to improve animal care and management and are willing to try it out. | For example, a sow gives birth to 10 piglets, and the bigger the pig, the faster they grow. The bigger the pig, the faster it grows. If the pigs don't grow well, they waste a lot of food, and it is not clear whether they will eventually become meat or not, so it is better to kill them from the beginning. But I still want to keep the pig because it has a life of its own. (Belief about the importance of human intervention, I02)<br>People who can see animals without smart technology can use smart technology effectively.<br>It's people like this who succeed. In short, if you are not interested in animals, smart technology is a waste of time and money, as I have seen. (Belief about the importance of human intervention, I04)<br>People who say, "Well, we are not doing well, so let's try to use the smart technology that is popular |
| | | | However, farmers who are not interested in animals often find smart technology to be a treasure. For farmers who have little interest in animals, technology adoption alone often does not solve the problem; farmers' interest in and intervention with their animals is important. Successful farmers have the ability to observe their animals closely and provide appropriate care. | now," are not doing well. For example, the success rate of artificial insemination is not good. Good farmers are able to see their cows properly, so the success rate is 80% to 90%. So, I think it is a fact that it is not good for bad farmers to remain bad, especially in the livestock industry. So, no matter how good the smart technology is, it will not work unless farmers are interested in animals. Farmers need to be interested in animals (Belief about the importance of human intervention, I04) |
| 44 | Introduction of smart technology | Deciding to introduce a new smart technology | Explore factors that influence this factor. A countervailing concept is the abandonment of introduction. | I wonder if they are actively adopting this kind of technology. (Access to technological information, I10)<br>I wonder if they will decide at that time whether they want to introduce this kind of technology or not. (I02_26)<br>I think that it is possible that the introduction of the technology will be decided soon. (I03_28)<br>I feel that it leads to the introduction of the system. (I06_17) |
| 45 | Investment capability | Farmers' investment margins and scale of operations that affect the ease of adopting smart technologies. | Farmers' investment margins and business scale are among the determinants of smart technology adoption. Large-scale farmers tend to have a large investment margin and actively introduce new technologies that are directly related to productivity and business effectiveness. Small-scale farmers, on the other hand, tend to have less investment leeway and value uniqueness and persistence. Farmers' economic status and income also affect the ease of adoption. High-income farmers are more likely to invest in new technologies, and farmers with higher production and farm income are more likely to adopt smart technologies.<br>On the other hand, if they do not have a managerial perspective, no amount of money will be spent on new capital investments, no matter how much financial leeway they have. | The price of each piece of equipment is not that high, so it is a matter of cost-effectiveness. (Perceived room for investment, I02)<br>I think there is no doubt about this. This also has to do with scale, and if you're large, you can afford it.<br>Large-scale companies can afford to invest in new technology if it directly affects productivity and business rather than being particular about it. Small-scale companies cannot afford to invest, and they value persistence and uniqueness. (Perceived room for investment, I03)<br>Pig farming is high income.<br>I have a high income, but I buy cars. They waste a lot of money. They buy cars from car dealers. This is my daughter's. This is my son's. This is my daughter's. This is my son's. So it would be better to raise their salaries and invest in production equipment. (Perceived room for investment, I06)<br>Whether or not the initial investment can be made<br>I think that the affordability of investment and financial support are linked. I think it is impossible to introduce a new technology if you cannot get financial support for it. (Perceived room for investment, I10)<br>The more farmers can afford the management and labor, the more they will introduce the technology (Production yield and income, I01)<br>There are so-so. Farms that are profitable (= farms that want to make money) are more likely to introduce the technology (Production yield and income, I02)<br>I think it is still easier to invest in technology when income is high. (Production yield and income, I03)<br>The higher the income, the more likely they are to work on it (Production yield and income, I04)<br>When I can afford it, or when I have a little bit of money to invest, I think I can make more money if I invest in this. So, I try to look for positive investment opportunities when I can afford them, and then invest in them to make more money. (I03_10) |
| 46 | Knowledge gap between farmers and agricultural cooperative officials | While farmers have many years of experience, the personnel in charge of agricultural cooperatives change regularly, resulting in differences in knowledge and experience. | Differences in knowledge and experience between farmers and their counterparts at agricultural cooperatives may influence the process of smart technology adoption by livestock farmers. If the personnel lack experience and knowledge, they may not be able to provide sufficient advice and support for the adoption and use of smart technologies. | Farmers are their own business, so if they are field farmers, they have been field farmers or potato farmers for decades, or if they are dairy farmers, they have been dairy farmers for decades, but the person in charge at the agricultural cooperative changes the person in charge after 3, 4 or 5 years, so they are not as good as a farmer forever, in terms of knowledge. (I08_261) |
| 47 | Lack of flexibility due to workload | Workload as a barrier or delay factor to technology adoption in the process of adopting smart technologies | Existing operations occupy many labor hours, making it difficult to devote sufficient time to addressing new technologies and operations. Labor shortages also contribute to barriers and delays in smart technology adoption. Understanding this and considering ways to reduce workload and workforce availability could lead to the widespread adoption of smart technologies. | One of the reasons for not doing trial and error is that the work is not mechanized, so there is no reason for it to be 8 hours a day, but if there are 8 working hours in a day, there is too much volume for one person to handle, and if there is existing work and new work to try, one factor is that the new work is not given time. One of the reasons for this is that there is too much volume for one person to handle. (Livestock breed, I06)<br>Small number of workers per farm (Livestock breed, I06) |
| 48 | Lack of investment capacity | Lack of certainty about economic constraints and risk avoidance when livestock farmers adopt smart technologies | Many livestock farmers are in dire straits and cannot afford new adventures. Although they have old and aging equipment, they are unable to afford to adopt new technologies even if subsidies are available. This lack of investment affordability is a barrier in the process of smart technology adoption by livestock farmers, and there is an attitude of risk aversion with the adoption of smart technology. In addition, non-best farmers have little or no investment leeway and cannot afford to adopt smart technologies when their production and farm income are high but they are still cycling through the process. These factors may slow the progression of the smart technology adoption process among livestock farmers. | I feel that many of the management bodies are at the very last level where failure is not an option. I think you need a certain amount of assurance to venture out.<br>There are many facilities that are 40 to 50 years old, and there are some that could benefit from subsidies, but it is difficult to take the plunge if you are not sure whether you can really succeed by introducing this technology. So, I think the reality is that there is not much time to renovate a part of the building, for example, and then change the whole building after confirming the results. (Entrepreneurship, I04)<br>Beef and dairy cattle farmers cannot afford it. (Perceived room for investment, I07)<br>Small-scale farmers can't afford to invest, and I think there is a place where they value persistence and uniqueness. (Perceived room for investment, I03)<br>I think that only the best farmers can afford it. (Perceived room for investment, I01)<br>Even if a farmer is making hundreds of millions of yen in sales, if he is running on a bicycle, he can't afford to put in the money (Production yield and income, I07)<br>Now, the price of feed is going up due to the problems in Russia and Ukraine and the exchange rate. So, farmers are in a situation where they can't afford to pay more. (I08_108) |
| 49 | Lack of psychological allowance | Degree to which willingness to cooperate with smart technologies is influenced by individual and situational mentality | There are individual differences in farmers' willingness to cooperate, and the degree of cooperation differs when psychological margins are present and when they are not. Generational change and personal circumstances are also factors that influence cooperation intentions, and these are related to psychological margins. The results suggest that an approach that takes psychological margins into account is important in the smart technology adoption process. | This may not be a very good example, but I asked a farmer with whom I had a brief relationship before to turn off the fan for one day so that the students could evaluate their stress afterwards, without even asking me. The farmer was young, and I knew him well, but I told him that I would not cooperate with the university if he said such a thing. (omission) There are people like that, and there are also people who are very open-minded and say, "Whatever you want, it's okay, try it out. It is true that each person is an individual. (I01_10)<br>I think we were just at the point of generational change, so we didn't have a lot of time to spare. When we asked the farmers who cooperated with us to cooperate with us again the following year, they said, "Sorry, we can't take care of them right now. So, well, maybe there is a reason for that. I think it was personal reasons for each farmer. I guess it depends on whether they have the mental capacity or not. (I01_12) |
| 50 | Lack of response to administrative support for subscription services | Livestock farmers are not receiving the subscription-based support they need from the government to continue to use smart technology | Government support for livestock farmers to adopt smart technologies focuses on lump sums and physical items. This suggests that the support diverges from where most smart technologies are subscription-based, and that there is not enough incentive for farmers to continue using smart technologies. Furthermore, policies are implemented on an annual | It's not so much that there's no support from the government, but rather, you know, you get a half subsidy or something, but in the end, [product name] or [product name] is something that's subscription-based and you're going to keep using it forever. So, instead of saying, "We built a barn, and it will be amortized over 35 years," you can say, "Well, how about this month, or next month? So, instead of saying, "I built a barn and amortized it over 35 years," they would say, "Well, how about this month or next month?" We can't subsidize subscriptions or things that are used continuously, so we have to make a 7-year contract. So, in the end, the only image we have is that of an investment in a mechanism, a box. (I08_285, 287, 299)<br>If you want to spread the software, then the software, you know, we subsidize 70%. So, if we were to offer a subsidy of 1,300 yen for the regular price of 2,000 yen to watch [name of video distribution |
| | | | basis, making it difficult to plan for the long term. | service], we would definitely get in. It would be better if the subsidies were half the price, but after all, government policies are made on a fiscal year basis, so it would be like saying, "We don't know what will happen next year, so let's give out the subsidies this year. (I08_303, 305) |
| 51 | Lack of sales and development manufacturers | Lack of manufacturers and low IT literacy of distributors preventing them from providing appropriate technology implementation assistance | Inadequate implementation support due to a lack of manufacturers and low IT literacy among agents is a key challenge in the smart technology adoption process for livestock farmers. | The number of manufacturers is small and the IT literacy of distributors is low, which makes it difficult to provide adequate support for farmers to adopt the technology (Others, I02) |
| 52 | Lack of transparency in individual management information | Farmers and managers who are not members of production associations tend to be reluctant to disclose data and results to the outside world. | In some cases, farmers who are not members of production associations are reluctant to introduce new technologies and share information. Management decisions may have a significant impact, and information exchange with other farmers may be limited. Therefore, a different approach may be needed for non-subscribing farmers. | As for places that have not done so (formed or joined a production association), I think it is probably up to the management and president. They don't really want to disclose data or their own performance to the public. They don't really want to show that they are doing smart livestock production. (Production degree centrality, I09) |
| 53 | Level of information disclosure in the industry | Impact of differences between business types where information disclosure is easy and those where disclosure is difficult | There are two types of livestock businesses: those in which information is relatively public, such as dairy farming and beef cattle, and those in which information is less public, such as pig farming, egg production, and poultry slaughtering. In those business categories where information is not well publicized, epidemics and grotesque details are relevant, and the distance from the general public makes it difficult to disclose information. This may affect the accessibility of information and technology. On the other hand, good farmers share information with the outside world by accepting outside inspections. Farmers who play a central role, such as the representative secretary of a producer association or the head of a youth club, are central to information disclosure and exchange. Increasing the level of information disclosure in the industry could lead to the diffusion of smart technologies and improved accessibility of technical information. | I think dairy farming is the most media-exposed type of livestock farming, especially milking. When the general public wants to become involved in livestock farming, I think they will first go to dairy farming. Beef cattle are also well publicized, but pig farming, egg production, and poultry farming are also very popular. Poultry farming is very closed, so people are not familiar with it, and in some cases, they don't know how it is actually produced. And there is a lot of grotesque content as well. It is difficult to disclose information. It is difficult to disclose information. Birds and pigs are vulnerable to diseases such as avian influenza, and it is difficult to disclose information to the general public because of the remoteness of the relationship between them. I wonder if there is any information that I can't find when I look for it. (Access to technological information, I06)<br>The better farmers accept outside inspection.<br>If you are the representative secretary of the producer's association, or the head of the youth club, etc. (Production degree centrality, I01) |
| 54 | Low interest in environmental considerations | Lack of interest in anything beyond environmental standards | Livestock farmers are focused on complying with legal environmental standards and avoiding complaints from neighbors. However, the emphasis is on super-efficiency, and interest in environmental impacts tends to be low. Regarding husbandry management, the results suggest that the emphasis is on | Not many producers are interested (Agri-environmental scheme membership, I01)<br>I have the image that pig farmers are all mainly focused on the environmental standards required by law and on not having their neighbors complain. They are not doing much more than that, are they? (Agri-environmental scheme membership, I02)<br>In Japan, the trend is toward hyper-efficiency, which is the opposite of the environment (Agri-environmental scheme membership, I03)<br>Breeding management may not be an influential factor.<br>As for breeding management, is it profitable or not? (Agri-environmental scheme membership, I04) |
| | | | profit.<br>On the other hand, there is a high interest in manure disposal, which can be an important motivator. | At least as far as manure disposal is concerned, I would expect them to be proactive (Agri-environmental scheme membership, I04)<br>(Chicken) Yes = environmental conservation = manure disposal<br>I think that environmental conservation includes the method of manure disposal. If money can be made for environmental conservation, and if this can be solved through smart agriculture, then I think that pig and poultry farmers will certainly participate in smart agriculture. (Agri-environmental scheme membership, I05)<br>I don't care...it's not the scale of management in terms of SDGs, ESG, etc., it's whether you can eat tomorrow, that's what's important, so everybody. (I08_33) |
| 55 | Negative buying experience | Past negative experiences with smart technology are a barrier to the next smart technology implementation | Past experiences with problems with equipment performance (measurement accuracy and durability) and doubts about reliability from manufacturers are disincentives for livestock farmers to adopt new smart technologies. | Since the equipment I bought before could not be utilized on the farm (poor measurement accuracy, poor durability), I assume that all similar equipment is no good.<br>The fact that the manufacturer might come out with something strange is also a disincentive (Others, I02).<br>So, when I first started this job, a very senior farmer who was very old and very old was very angry with me, saying, "People like you come out when the price of cows is high, but you disappear as soon as the price of cows gets bad. He was very angry with me, saying, "We still have to do what we do. So, Smart Farming itself has a long history of [company name] and [company name] coming in, doing things properly, and then pulling them out, and then pulling them back in, and I think there is a lack of trust in Smart Farming. (I08_195, 197, 199) |
| 56 | New entrants from other industries | People new to the livestock industry or moving from other industries who are interested in smart technology | Newcomers and those who have moved from other industries may be more interested in environmental conservation and public benefits and more open to new technologies, making them more likely to adopt smart technologies than existing farmers. | Mid-career hiring. People who have been in other types of businesses are more likely to be introduced (Socio-demographic characteristics, I10)<br>People who have not kept cattle before and who want to keep cattle are deeply interested in environmental conservation and public interest in the market, and are willing to do so. (Agri-environmental scheme membership, I07)<br>Newcomers want to do smart farming, but existing farmers may have a hard time getting in.<br>Young people who have left the farm to start their own business, for example, all say they want to do smart farming, but existing farmers have a hard time getting into it. It is difficult for existing farmers to enter the market. (Others, I07) |
| 57 | Paradigm shift to animal welfare | Shift in values and mindset to recognize the importance of animal welfare in the livestock industry and work together with environmental conservation and productivity improvement | The adoption of smart technology is not just about profit-seeking and efficiency, but also contributes to animal welfare and environmental conservation. For livestock farmers, paradigm shift is an important process to change the vector of the country as a whole and agriculture as a whole, and achieving this may lead to more widespread adoption of smart technologies and contribute to overall agricultural sustainability.<br>On the other hand, it is suggested that few farmers are willing to adopt smart technologies while considering the paradigm shift, which requires action at the national policy level. | It's an increase in productivity, but we have to improve productivity on top of such animal welfare, circulation, and environmental protection systems. The most important motivation is to make money and make life easier, but this shift in the system is the vector for the future of agriculture as a whole, and the country as a whole needs a paradigm shift. So, I don't think it's a matter of putting smart things into the system. What do you think? (I07_8) |
| 58 | Perception of low risk | Situations where you feel there is minimal risk or burden in adopting a new smart technology | The results suggest that when livestock farmers adopt smart technologies, it is important for them to be cautious in making decisions based on the successes of others, to feel secure when financial assurances are available, and to be in a | I wonder if many people don't want to be the first penguin unless they see the effect of someone else's work. Hmmm. (I04_10)<br>In order to have the farmers prove their work, we write a contract that guarantees a profit or loss, and although we don't do this very often, we do ask the farmers to prove their work while we write the contract. It is difficult to know what the trigger is for such farmers, though. Hmmm... I think they feel that if they don't lose anything in the trial, they will let us try it anyway. Hmmm... (I04_14) |
| | | | situation where the budgetary burden is low. With these factors in place, livestock farmers may be more willing to experiment with new technologies. | I guess it means that there is no or little budget burden. So, if there is a subsidy or joint research with a university, or if they don't have to bear the cost of equipment, then they can be included in the program as long as the cost is small. (I07_14) |
| 59 | Positive experience from existing systems | Useful experience and knowledge gained from using existing smart technologies and systems | It is suggested that livestock farmers are using existing smart technologies such as milking systems and herd management software and are gaining positive experiences from them. Positive experiences from existing systems may influence adoption and adaptation of new technologies, regardless of generation. | There are many people in their 60s and 70s who are using the system. As I mentioned in the management section, dairy farmers have already introduced milking systems. And those who are using it well are making good use of the herd management software called V-Map Three, which is produced by [company name]. In such places, the older generation is already familiar with the software. I wonder if there might be some kind of relationship with the breed of cattle in such places. (I01_2) |
| 60 | Pressure to comply with animal welfare standards | Pressure exerted by those who value animal welfare | Pressure is sometimes applied to the livestock industry by those who value animal welfare. However, this pressure is not pressure to adopt smart technology, but pressure regarding the care and handling of livestock. For example, there may be pressure to adopt smart technology in order to practice the husbandry practices demanded by major retailers, but pressure for animal welfare is not necessarily equated with the adoption of smart technology. | (Who are they yelling from?) First of all, the person who wants to put the animal on animal welfare or something like that is going to attack us. (Access to technological information, I06)<br>The pressure now is the pressure for animal welfare, not the pressure to introduce smart technology. I don't think it's a question of judging smart technology by that pressure. It's not that, it's the way we keep them, and Costco wants us to do it, so we do it. Because it is not equal smart technology. (External pressure, I05) |
| 61 | Product reliability | When considering the implementation of smart technologies, attitudes that emphasize best practices and the reproducibility of the quantitative effects of the technology. | Livestock farmers are more likely to adopt a technology if they can confirm case studies and quantitative reproducibility of effects when introducing the technology. It is easier to introduce a technology when successful case studies are presented, and information from experts such as feed companies, technology extension agents, and veterinarians is also important. However, livestock farmers often want to hear not only information from experts, but also the experiences of other farmers who have actually introduced the technology to confirm its effectiveness. | If case studies and quantitative reproducibility can be confirmed, adoption is more likely (Perceived usefulness, I06)<br>It is easier to adopt if there is an introduction of successful case studies.<br>Feed companies can basically introduce such cases. Like technology extension agents. Or veterinarians.<br>But even if we hear about it from such people, we still have to ask ourselves, "Is it really effective? Are there any farmers who are doing it? Then, let's hear what the farmer has to say and introduce it. Then, I'll ask the farmer if he wants to talk about it, and introduce it. (Support institutes/programs, I09)<br>I think that the trust in the sensor is not so great, and I think that ABS was like that at first. (I08_166)<br>Also, as for pressure from outside, I think that word of mouth is important, for example, "That farmer here has improved his business by introducing this," or "He is now able to produce good meat. (I09_39) |
| 62 | Promoting the entry of young people | Efforts to increase the attractiveness of livestock farming by introducing new technologies to make it easier for the younger generation to enter the livestock farming industry | It is important for livestock farmers to showcase innovation and progress by adopting new technologies to attract the next generation of young people. This process could serve as part of the motivation for livestock farmers to adopt smart technologies, and could also have the effect of improving the competitiveness of the industry as a whole by encouraging the entry of younger generations. | Also, for example, if there is an increase in the number of young people, or if a company is raising pigs and wants to involve young people in the process of replacing them, or wants to hire more people, one factor that could attract such people is to show them that the company has introduced such technology and is challenging new things. It may be a good opportunity to show them that we are trying new things by introducing this kind of technology. (I09_12) |
| 63 | Psychological allowance | Time and money availability, open-mindedness and risk-taking in adopting new technologies. | It has been suggested that a larger farm size creates a psychological margin in the smart technology adoption process for livestock farmers. This psychological leeway manifests itself as time and financial leeway, and may be a factor that encourages positive attitudes toward technology adoption and experimentation. | On the other hand, larger companies have a little more leeway, so they can introduce technologies that allow them to do things without hiring people or spending time and money. I have the image that the factors that make such a difference in time and money vary depending on the scale of the project. (I03_20) |
| 64 | Recognition of economic benefits | Understanding of the economic benefits of adopting smart technologies, such as increased production efficiency, cost savings, and return on investment (ROI) | One of the deciding factors in whether livestock farmers will adopt smart technologies is their perceived economic benefits. Livestock farmers may be more willing to adopt smart technology if they can realize tangible benefits and cost savings or if the return on investment is clear. In addition, young livestock farmers have an entrepreneurial spirit to improve production efficiency and save labor, and such thinking may influence their decision to adopt smart technologies.<br>Technology awareness and support are important to improve such awareness.<br>The link to the challenge of new business and business expansion is particularly strong among younger workers.<br>In the absence of a managerial perspective, they may not fully recognize the economic benefits. | It's not easy if you don't realize things like cost savings from disease detection, or profit that comes out in numbers. I would say if there is an increase in profit when you actually look at the books for the year (Perceived usefulness, I04).<br>I would accept it if it is an important point related to production efficiency. (Belief about the importance of human intervention, I07)<br>Can you make a return on invested capital?<br>Can you understand the ROI of smart technology (Perceived room for investment, I08)<br>Some young people are aiming to expand the number of their employees, so they are more interested in improving production efficiency, including labor saving, of course (Entrepreneurship, I07)<br>I think that the final goal is to increase sales, or what is it, to make management easier, but I think it is to increase near-equal sales. (I02_87)<br>I think the most important thing is that this technology seems to be profitable. (I03_2)<br>IoT itself is productive and labor-saving, so if it is known that it can pay for itself enough to cover the cost of the technology, even if it has to borrow money, then I think it can only be a good thing. (I03_48)<br>Rather than specific customers, the way we sell, the way we sell, the way we complain, is that if a woman misses her first chance of conception, she is forced to eat for a month, and since the ovulation cycle is a 21-day cycle, we have always said that it is meaningless to force a woman to eat for 21 days. So, if we put 40 days as the period after calving and after the uterus recovers, well, if we can get the calf pregnant in the next heat, we can have many calves in the shortest cycle, and I think that whether you can understand that or not, we can gain management profit by shortening that period. So, people who know that the profit they can gain is large in relation to the cost of installing sensors, etc., will do it right away. But the reason why people don't understand is that they don't understand the management skills and business conditions. (I08_174, 175, 177)<br>I think sales will go up. (I09_2) |
| 65 | Reflection of efforts | Efforts and initiatives made by livestock farmers are directly reflected in product prices and market valuation | Livestock farmers want their efforts to be reflected in product prices and market valuations. Through the reflection of their efforts, farmers can communicate value to consumers, and consumers are likely to pay higher prices for products with high quality and uniqueness. This effect is expected to improve farmers' profitability and stabilize their operations. To encourage reflection of effort, it is important to provide information and marketing that enables consumers to recognize the quality and characteristics of products. An environment that reflects effort may also increase the sense of competition among farmers and contribute | Perhaps something that reflects effort. For example, good milk can be purchased for 400 yen per liter, and it would be good if there was a mechanism to make a difference of 2, 5, or 10 times rather than 10 per cent. For example, there is an area in Hokkaido called [name of company] or [name of milk product] that produces only milk with low lactose content for people who get sick to their stomachs from the presence of lactose. There are regions that produce only low-lactose milk for people who have stomach problems due to the presence of lactose, and there are regions that add value by putting out the [product name for milk] and selling it at twice the price of regular milk, or selling it as organic milk. I think it would be better if everyone started to look at management. I think the way to do this should probably be more sensitive, but since agriculture is a key industry of the country and must be protected, I wonder if this is the right way to go about it. (I08_235, 237, 239, 247) |
| | | | to improving the quality of the industry as a whole. | |
| 66 | Responce to market trends | Be aware of market trends and change rearing methods, suppliers, shipping methods, etc. according to such changes. | Livestock producers need to be constantly aware of market trends in order to select suppliers and shipping methods. For example, in the poultry industry, they may focus on doing business with large foreign companies. Livestock producers may also use diversified shipping methods, such as shipping at market prices or using animal welfare (AW) rearing. Furthermore, they may increase the number of animals raised depending on market trends. For example, a spike in pork prices in the pork market may trigger an increase in the number of animals raised. Responding to market trends is important for livestock producers to remain competitive and ensure business sustainability. | (Chickens) Very aware of business with foreign companies (Costco, McDonald's) (Market knowledge, I05)<br>(Chickens) They ship chickens at market price, and on the other hand, they ship chickens to AW farms, and they are diversifying their business. (Market knowledge, I05)<br>Trends in the pork market. For example, a high hog price will trigger an increase in the number of pigs (Perceived room for investment, I09) |
| 67 | Restrictions on information exchange | Difficulty for livestock farmers to exchange information or for manufacturers and developers to properly inform producers in terms of quarantine, etc. | Quarantine issues and business characteristics may limit the exchange of information among livestock farmers. In addition, technology developers may have difficulty entering farms to communicate information due to epidemics, so digital technology that is easy to use and results easy to understand is required. This limitation in information exchange has slowed the diffusion of smart technologies. | In the case of pig farming, it is word of mouth. Generally speaking, there are many rumors about this type of business, but because of the quarantine aspect of hog cholera, it is not easy to get together and exchange information this time. Manufacturers also want to communicate information, but they are not able to give good information to the producers. (Access to technological information, I06)<br>The people who develop digital tools for pig farming materials have difficulty communicating how to use them on the farm due to the epidemic. Without digital technology, there would be misunderstandings and the results would not be what you were looking for, which would slow down the spread. (External pressure, I06) |
| 68 | Risk hedge | To use digital technology to hedge risks in terms of quarantine and farm size and location. | There are cases where bases for quarantine are dispersed. In some cases, digital technology is used to improve efficiency. In addition to quarantine, there are also cases where bases are dispersed due to land use restrictions, and digital technology plays an important role in these cases as well. | Due to biosecurity concerns, there is a tendency to decentralize across multiple locations (risk). In other words, even with communication lines installed, operating two locations as if they were one is where digital excels, bridging gaps, which I believe is a factor for its adoption. For pigs and poultry, species that are particularly susceptible to disease, there are cases where risk hedging is done by decentralizing locations. People often mention Japan's agricultural productivity, but there is an issue with securing vast farmlands, as is the case with rice paddies. By decentralizing and thus mitigating risks, addressing this digitally could lead to improved productivity. In any case, even on a farm, it can take an hour or 30 minutes to walk from one end to the other. Digital solutions are crucial for such scenarios, such as digitizing movement or monitoring via screens so that people can come from the other side when needed. I believe this is the most necessary approach. (Farm size, I06) |
| 69 | Self-efficacy of technology use | Refers to the confidence or belief that users can operate and utilize smart technology | In the smart technology adoption process, self-efficacy for technology use plays an important role in the decision to adopt and use technology. The results suggest that there are differences in self-efficacy depending on the user's age and technological literacy. Users with high self-efficacy tend to be more proactive toward new technologies, which may facilitate technology adoption and use. | For example, if they can use LINE on their smartphones, they can use it if they have a smartphone, not if they have a cellular phone. I think it's about giving examples of users. (I02_38) |
| 70 | Sense of competition with neighboring farmers | Livestock farmers' competitive attitude in terms of milk quality and production efficiency compared to neighboring farmers. | Some livestock farmers have a competitive mindset with surrounding farmers and focus on milk quality and production efficiency in their daily management. This competitive mindset is a motivating force for farmers in the region, driving them to achieve good milk quality and efficient production, but it also puts pressure on them at the same time. It is important to consider livestock farmers' sense of competition with neighboring farmers in the diffusion and development of smart technologies.<br>A managerial perspective can lead to a sense of competition. | Also, milk quality is often ranked as the best in the region. Good farmers are very concerned about this, and they use it as motivation to manage their farms on a daily basis. But in a sense, motivation is a kind of "Kamihitoe," and I think it can become a source of pressure. So, I think that farmers who do not care too much about milk quantity and quality, but work hard, and who will be responsible for the region in the future, are very concerned about it, and that is why I think that such farmers are important. (Perceived usefulness, I05)<br>I think they will ask if there is a clear indication of profit (expansion of company scale, etc.). (External pressure, I03)<br>The first is to compare their performance with other farmers, to compare their performance with other farmers, and to see if their performance is actually not high (I02_56).<br>The first is to collect production information from farmers all over the country at once, and then benchmark their performance, such as the highest performance of this person and the lowest performance of that person. (I02_67) |
| 71 | Shift in values due to technological adoption | Impact of smart technology adoption on livestock farmers' values and management style. | By implementing smart technology, traditional labor can be reduced and the time can be used to improve other management skills and monitor business conditions. This could result in a process that encourages further smart | On the other hand, some customers who have introduced the system say, "I really don't have to do heat detection anymore, ha ha ha. I think that this is a trend toward smarter and smarter management, whereby management skills can be refined and management conditions can be monitored. (I08_168, 170) |
| 72 | Small-scale and niche breeding methods | Small-scale, niche breeding practices that are not mainstream influence the adoption and adaptability of smart technologies | The diversity of husbandry practices influences the diffusion of smart technologies. Where common and mainstream rearing practices are employed, smart technologies are easy to adopt. However, for farmers who employ unique husbandry methods, appropriate smart technologies may not exist, making adoption difficult. In addition, farmers who focus on the capabilities of individual animals may delay the introduction of smart technology because there are problems that cannot be solved by smart technology. Thus, the diversity of husbandry practices affects the diffusion process of smart technology. | Somewhat likely. Not many products for farmers with niche husbandry practices.<br>It would be very easy to spread if livestock keeping methods were standardized across the country Farmers who have a general mainstream way of keeping livestock are easy to introduce. So there are farms that have unique ways of keeping livestock and so on. In the pig industry, freestalls are in the minority, so there are not many products suitable for freestalls (Management type, I02).<br>On the other hand, farmers who aim for high capacity per milking cow are not likely to rely on smart technology, because the management of each cow requires a very high level of technology.<br>The point is that when chasing something like individual merit, rather than economies of scale, there is a nuance that says there are problems that smart technology cannot yet solve. (Production yield and income, I04)<br>I'm talking about chicken farmers, and the majority of them are small-scale farmers who don't really have the situation to put in IoT or anything shiny, I don't know what it is, but I tend to look at it (I03_20) |
| 73 | Succession of business to younger generations | To pass on the business conducted by the parents' generation to the children's generation to continue running the business. | Business succession has become an important issue in the livestock industry. In some cases, young people are entering the industry, and management is being passed on from the parents' generation to their children's generation. In many cases, the next generation is becoming the next generation in business management and possesses advanced knowledge. In addition, when income is abundant, the younger generation is more likely to take on challenges. These factors are thought to influence the process of business | The pig industry is now very much in the process of being passed on to younger generations. (Access to technological information, I06)<br>After all, it was after the war of the first generation. The second and third generation of livestock farmers came back from the war and said they had no land. The second and third generations had no land, so they cleared the mountains and moved into the livestock industry. At that time, due to the U.S. grain policy, the Japanese market was chosen to process the bumper corn and wheat crop, and the livestock industry developed along with the reconstruction of Japan. I think there is a tendency that the Japanese market was chosen for the processing of corn wheat, and the livestock industry developed along with the reconstruction of Japan,<br>The first generation had no money, but the pigs were getting bigger and bigger. The first generation had no money, but they kept increasing the number of pigs. They did it in their own way, but in order to pass it on to their children, they sent their children to XXXX or to study abroad. (Market knowledge, I06) |
| | | | succession<br>Younger generations tend to use IT to gather information and introduce new technologies. | The next generation is on the way. If they can't pass on their business, they will go bankrupt.<br>I guess it's the income I mentioned earlier. I think that the younger generation will be able to take on new challenges because there is a certain amount of leeway in the "income" category. (Socio-demographic characteristics, I06)<br>There seems to be an influence of generational change (External pressure, I08)<br>In terms of private management, I think that generational change is very important in a family business. (I09_38)<br>I think that the point where people change is important. It's about generational change. (I10_14) |
| 74 | Support organizations and programs | Institutions and programs that serve to effectively introduce new technologies to farmers and support real-world problem solving. | It has been noted that prefectural extension agents and agricultural cooperatives play an important role in technology dissemination, but they are aggressive in favor of technologies that they can benefit from. On the other hand, even when good technologies developed by research institutes are presented, it is sometimes felt that support programs are lacking in solving problems in each field and in actual use. Thus, the lack of sufficient human resources and time in technology dissemination and problem solving has become an issue. | Are Prefectural Extension Agents Important for Technology Dissemination? Agricultural cooperatives are proactive about technology that they can sell and make money for themselves (Support institutes/programs, I04)<br>It's fine for research institutes to make good technology and present it in a bong. However, when a problem arises in the actual use of the technology, it is not easy to say, "Let's do this," "Let's do that," or "Let's do that. Without a support program, it would be very difficult without someone who has the skills and farmers. Of course, the more support we have, the better, but the problem is that we don't have enough people and time. It is difficult to say whether we are doing enough. (Support institutes/programs, I03) |
| 75 | Technological adoption barriers for women | Challenges and hurdles associated with the adoption of smart technologies by women livestock farmers, as well as requirements and support for women to become proficient in the use of technology | A key factor in the adoption of smart technology is whether it can be handled by women. Since half of the labor force in family-owned livestock farms consists of women, the technology must be provided in a form that is easy for women to use. It is also important to approach women in publicizing the technology, and such efforts may have the effect of lowering the hurdle. | What women cannot handle is that men like to ride on tractors and other machines. So, when it comes to smart agriculture, I think the current generation can operate various types of machines without much resistance. But on the other hand, I think women need to be able to use these machines as well. I have seen in real life that even milking robots can't be used by the wife because she can't handle ...... and the husband can't handle it when he goes on a business trip, and so on. I think the complexity of women not being able to handle them is a huge hurdle. (I05_51, 52, 54)<br>Also, I sometimes see women working on tractors nowadays, but I think that tractors are a man's job. If there is a PR campaign that shows that a woman can solve a problem, or that she can be kind to others, the hurdle will be lowered and people will think, "Oh, my wife can do that, too. So, I think that many machines are marketed to men, but in the end, it is women who do the hard work of helping others, so I think it would be good if there was a way to make it easy for women, too. (I05_60, 64)<br>Even though it is a family business, the number of cows is still large, so I would like to introduce this technology effectively as a technique that cannot be fully covered by the family. Half of the workforce is made up of women, so I hope that this point of view will not be ignored. (I05_68) |
| 76 | Technology education and support | The role that technology-related assistance plays in the process of farmers' adoption of smart technologies. | Technology awareness and technology adoption support is important to ensure that livestock producers are able to effectively use new technologies and devices. This requires not only teaching them how to use the devices, but also providing guidance on how their role can lead to increased productivity. Livestock producers ultimately want to make sales, and it is important that they understand how new technologies can contribute to sales. The government and manufacturers are expected to play a role in this activity. | (Farmers) follow the information provided by the JA (Japan Agricultural Cooperatives) and government. (Market knowledge, I07)<br>Technical support (awareness of operations leading to results) is crucial.<br>In the case of pigs, about 30% of the feed supply comes from agricultural cooperatives, while the rest is from private companies. Since feed costs account for about half of the total costs, the relationship with feed suppliers is extremely important. As they hold half of the costs in pig farming, agricultural cooperatives have mechanisms such as advisors who provide guidance. (Support institutes/programs, I06)<br>I think this is also important. Often, things are just created and then left as they are.<br>It is important to instruct on how devices can lead to improved productivity.<br>Manufacturers understand and can explain better than government officials.<br>If the role of the device is not clearly shown to be linked to productivity improvements, it is meaningless. If it's just about teaching how to use it for free, farmers can learn it quickly. |
| | | | In particular, manufacturers are expected to be more knowledgeable about devices and technologies than the government and to be able to communicate more smoothly with livestock producers. In addition, companies that sell products such as feed and smart technology could actively provide technical support and dissemination activities to promote the effectiveness and benefits of the technology. | But the real issue is whether they continue to use it. The significance of smart devices lies in this. Since farmers' ultimate goal is to increase their revenue, it's important to explain how using the device will lead to increased sales. Without clear communication and support, it will be difficult to achieve this. Therefore, knowledgeable support personnel are essential. Compared to government officials, I believe manufacturers understand this aspect better. (Support institutes/programs, I10)<br>Even with feed, the motivation to promote it comes from the desire to sell the product by claiming it will increase milk production or improve the cow's body condition. Similarly, companies selling smart technologies will provide guidance. Companies selling technology, like DeLaval or Orion in dairy farming, would be more inclined to do this. (Support institutes/programs, I04)<br>For products like [product name], where it is hard to explain verbally how to ride a bicycle or tie a shoelace, proper on-site demonstration and guidance on correct operation lead to recognizing the correct effects and result in adoption. (I06_17) |
| 77 | Tendency to maintain the status quo | Tendency to be reluctant to adopt smart technologies and maintain current operations and production methods | If the business is small, such as a family business, and the owner is satisfied with the status quo or has no desire to make much money, there is a pattern of not introducing technology. Also, there may be a decision to maintain the status quo (not to do anything new) because the business is running well.<br>In some cases, the lack of a managerial perspective may work to maintain the status quo. | Those who want to maintain the status quo and live with pigs leisurely are not likely to introduce the system (Production yield and income, I02).<br>Production yields are not so high in places where people are not very conscious of making a profit. (Production yield and income, I02)<br>Places where management is in a good pattern don't have to accept it anymore. I mean, those places have already put in some IoT or something. Some places that are running well don't need it anymore. (Production yield and income, I07)<br>So, people who are fine with the status quo, who are not troubled by the status quo, and who do not want to change the current way of doing things are not going to make much progress in the adoption of IoT. (I02_89)<br>I have the image that they are not willing to adopt new technology or knowledge because they can do it now. (I03_8)<br>I think that there are various compounding factors, and that they think that they can make a living by doing their daily work, so they are not really interested in running a business, and if they have any problems, they just go to the agricultural cooperative and yell at them. (I08_201) |
| 78 | Time lag in information access | Time delays in farmers' access to agricultural technical information. | If it takes time for farmers to access information, there may be a delay before information disseminated by the information provider is actually applied in the field. It is important for information providers to provide information in a format that is easy for recipients to access and to reduce the access time lag.<br><br>Slowness in information transmission is affected by the type of livestock. | Longer time to access to information (Access to technological information, I06) |
| 79 | Time lag to realize value | The time it takes for the effects and benefits to be realized after the introduction of smart technology. | The time lag to realize value is an important factor for livestock farmers when considering the adoption or continuation of smart technology. Smart technology may not be a "must-have" because some functions can be performed using conventional methods. The decision to adopt or continue is influenced by the fact that smart technologies do not always provide immediate benefits and some functions can be performed using conventional methods. | Hmm, first of all, it takes time to realize the value. For example, detecting estrus or identifying diseases can be done by people working on the farm, so it's not a must-have. So, when the market conditions worsen, what happens is that even if they cancel the service, farm management can still continue, right? It's not a must-have product. So, for example, when plows pulled by cows or horses were replaced by tractors, that was a must-have innovation. I think there was innovation happening there. But what [product name] is doing doesn't seem to be bringing that level of innovation to the industry. (I08_144, 146, 148) |
| 80 | Tradeoffs in technology adoption | A phenomenon in which the advantages of adopting smart technology are countered by disadvantages in terms of quality, flavor, etc. | When livestock farmers adopt smart technologies, they consider the trade-off between the advantages and disadvantages that the technology brings. In the case of the milking robot, while it saves labor, it has the disadvantage of reducing the flavor of milk. These trade-offs must be understood, and decisions must be made to adopt the appropriate technology for each farmer. It is important for technology developers and extension staff to clearly communicate the advantages and disadvantages for livestock farmers and support them in making appropriate choices. | Actually, milking robots can measure in real time. However, research has shown that milking with a milking robot increases rancidity, a substance in milk quality that induces abnormal flavor. Therefore, it is not desirable for farmers to have milking robots that make milking easier but increase Rancid in the milk quality and deteriorate the flavor of the milk. (Perceived usefulness, I05) |
| 81 | Traditional view of agriculture | A mindset that is resistant to adopting new technologies and adhering to past methods and practices | The traditional view of agriculture may be a factor or barrier that delays the adoption of smart technology. This view suggests that there is limited access to new technology and knowledge and adherence to past successes. | Farmers are already talking like people should watch their cows. (I08_83)<br>It's similar to what I said earlier about the parent-child relationship, but I don't know anything else. I have the image of a person who thinks that riding in a manual go-cart-like car without ABS or automatic transmission is the only way to drive a car. (I08_85) |
| 82 | Transitional period in animal care | Transitional period with the need to change breeding methods and production systems to meet market trends and animal welfare requirements | The livestock industry needs to change its rearing methods and production systems to meet the needs of markets and consumers. In particular, with the growing demand for animal welfare, conventional methods of animal husbandry may no longer work. On the other hand, without profitable mechanisms and subsidies, it is difficult to change husbandry methods. Government support is required during the transitional period. | (Chickens) They are shipping at market price, and conversely, shipping to AW feeders, and diversifying.<br>(Chickens) They are quite conscious of doing business with foreign companies (Costco, McDonald's) (Market knowledge, I05)<br>However, if animal welfare is introduced for both pigs and chickens in the next 10 years or so, the way they are kept will be different, and the concept will be completely different again. I think it will change completely. For example, there are cows that are chained together and live on one tatami mat for the rest of their lives. They stand up, sit down, get milked, eat food, and urinate and defecate. There are also Wagyu cows that are chained together. It is no longer possible to do that now, so it would be better if the cows were chained together, because it would be easier for the humans to manage them because the area to be managed is smaller. But that kind of thing will no longer be possible. Even chickens are kept in cages, one by one. They fight with each other. Chickens live in a harem, so they fight. Chickens are kept in cages to prevent this and improve the performance of each individual chicken, but this is no longer acceptable, and Europe is doing the same thing. But if they are allowed to roam free, their productivity is reduced. And the weak ones are bullied and die. Eggs are laid here and there, and there is feces everywhere. And the humans have to pick them up. That's the way to keep them. It's very systematic now, for the convenience of humans. I think this will change in 10 to 15 years. Then the system will be totally different. However, I have heard that MAFF is planning to use its budget for animal welfare, so they will probably try to subsidize the transition during the transitional period.<br>I think that you can't use the IoT without first creating a profitable system. (from Livestock breed, I07)<br>However, as I mentioned earlier, we are now at a turning point, and the human-centered way of raising animals that suited the high-growth period up to now will no longer be possible. Chickens and pigs will also become more difficult. (Perceived room for investment, I07) |
| 83 | Understanding of family | In family businesses, it is easier to implement smart technology when there is family understanding | In a family business, family understanding is an important factor in technology adoption. When family members understand the benefits of technology, | As far as dairy farming is concerned, it's very much a family business. People in the immediate area are involved in the work. So it is easy to introduce dairy farming not only for their own labor force, but also to reduce the labor force of the people in the close circle. I think it would be easier to introduce dairy farms if the family has a good understanding of the dairy farmer's needs. (Others, |
| | | | labor savings, and managerial advantages, technology adoption is likely to proceed smoothly. | I05)<br>In a family business, for example, if the mother is doing the breeding and the father is doing the artisanal breeding, how should I put it, the father's voice is strong. If the father does not share that feeling, I don't think it will be introduced. (I09_28) |
| 84 | Withdrawal of introduction | Considered implementing smart technology but decided against it. | Livestock farmers may begin to consider smart technology adoption but make a last minute decision not to implement it. A counterpart concept is the introduction of smart technology. | We often hear stories of adoption being abandoned due to price issues. (Perceived room for investment, I02)<br>I don't think it will be introduced. (I09_28) |